\documentclass[letterpaper,twocolumn,10pt]{article}
\usepackage{usenix-2020-09}
\usepackage{authblk}
\usepackage{todonotes}
\usepackage[utf8]{inputenc}
\usepackage[T1]{fontenc}
\usepackage{graphicx}
\usepackage{grffile}
\usepackage{longtable}
\usepackage{wrapfig}
\usepackage{rotating}
\usepackage[normalem]{ulem}
\usepackage{amsmath}
\usepackage{textcomp}

\usepackage{amssymb}
\usepackage{capt-of}
\usepackage{hyperref}
\usepackage{tikz}
\usepackage{mathtools}
\usepackage{array}
\usepackage{flexisym}
\usepackage{subcaption}
\usepackage{paralist}
\usepackage{algorithm}
\usepackage[noend]{algpseudocode}
\usepackage{pgfplots}
\usepackage[title]{appendix}
\usepackage{xspace}
\usepackage[english]{babel}
\usepackage{blindtext}
\usepackage{graphicx}
\usepackage{makecell}
\usepackage{comment}

\newcolumntype{C}[1]{>{\centering\arraybackslash}m{#1}}
\newcolumntype{R}[1]{>{\raggedleft\arraybackslash}m{#1}}
\newcolumntype{L}[1]{>{\raggedright\arraybackslash}m{#1}}
\usepackage{multirow}

\newcommand{\squishlist}{
  \begin{list}{$\bullet$}{
    \setlength{\itemsep}{0pt}       \setlength{\parsep}{3pt}
    \setlength{\topsep}{3pt}        \setlength{\partopsep}{0pt}
    \setlength{\leftmargin}{1em}    \setlength{\labelwidth}{1em}
    \setlength{\labelsep}{0.5em} } }

\newcommand{\squishend}{
  \end{list} }

\usepackage[small,compact]{titlesec}
\usepackage{amssymb}
\usepackage{pifont}

\usepackage{stmaryrd}

\usepackage{algorithm}
\usepackage[noend]{algpseudocode}
\usepackage{algpseudocode,color,colortbl}

\RequirePackage[normalem]{ulem}
\RequirePackage{color}\definecolor{RED}{rgb}{1,0,0}\definecolor{BLUE}{rgb}{0,0,1}

\begin{document}

\newcommand{\ming}[1]{}
\newcommand{\zerui}[1]{}

\date{}

\title{Building A CSFQ-Inspired Transport for Switched CXL Memory Pooling}

\makeatletter 
\renewcommand\AB@affilsepx{\hspace{2mm} \protect\Affilfont} 
\makeatother

\author[1]{Zerui Guo}
\author[2]{Emily Shriver}
\author[1]{Ming Liu}
\affil[1]{University of Wisconsin-Madison}
\affil[2]{Intel}

\newcommand{\sys}{MemChannel\xspace}

\pagenumbering{gobble}

\maketitle
\begin{abstract}
\label{sec:abst}

Emerging switched CXL memory pooling systems, albeit promising, suffer from significant performance interference due to the shared but performance-uncontrolled data path among concurrent memory streams between a host core and a remote DIMM. We systematically characterize a memory pooling appliance based on the XConn’s Apollo CXL switch, and identify three issues: intra-host contention, in-fabric congestion, and unmanaged host-remote DIMM interaction.


This paper presents a new transport layer, {\bf \sys}, which provides the \texttt{mchannel} abstraction to manage end-to-end fabric bandwidth among competing memory flows and enable application-specific traffic for switched CXL memory pooling. Under the hood, our key idea is to build a {\it Sender-Driven Fabric-Informed} transport protocol--inspired by Core-Stateless Fair Queueing (CSFQ)--that admits just the right amount of CXL requests to each \texttt{mchannel} based on the estimated $Core \leftrightarrow DIMM_{CXL}$ bandwidth availability. To grapple with the ramifications of CXL-induced idiosyncrasies, \sys introduces a couple of techniques: time-based rate control, host-side admission control, cross-host bookkeeping, new congestion signals, rate estimation based on the fluid model, and delay-based link capacity adjustment. We build \sys from scratch and support unmodified applications. Our evaluations over switched memory pooling demonstrate the effectiveness of \sys from performance isolation, scalability, and multi-tenancy perspectives.


\end{abstract}

\section{Introduction}
\label{sec:intro}

Load-Store interconnects--such as Compute Express Link (CXL)--and the resulting memory pooling have gained significant traction recently. They transparently extend a commodity server's memory capacity, enable elastic memory resource sharing, and allow applications to access remote memory using load/store instructions. Such composable memory technology holds great potential to remedy the DRAM scaling issue, ameliorate hardware resource utilization, and boost the cost efficiency of rack/cluster-scale computing. The last few years have seen several industrial prototypes~\cite{asteralab-leo, omega-fabric, intel-agilex-dev-kit, xconn-titan, samsung-cmmb} being developed, evaluated, and sampled.

However, memory pooling, especially under switched deployment, suffers from significant performance interference. Unlike local memory connected via dedicated memory buses, {\it the data path between a host core and a remote CXL DIMM under switched pooling is shared among concurrent intra-/inter-host memory streams without explicit performance control.} We systematically characterize the problem using the XConn's Apollo CXL switch and Titan evaluation platform and identify three issues ($\S$\ref{subsec:chara-issues}). First, at the server host, taking the Intel EMR (Emerald Rapids) processor as an example, access contentions happen at both (a) the CHA (caching and home agent) and M2PCIe modules of the host uncore; and (b) the virtual lane inside a CXL host adapter, increasing the CXL memory access latency by up to 13.4$\times$ with 82.6\% bandwidth drops. Second, within the switch fabric, the link-layer credit-based flow control is workload-agnostic, causing head-of-link blocking and unfair bandwidth partition at the egress port. As such, when a cacheline-sized memory stream interleaves with a 4KB-sized one, it experiences 6.9$\times$/3.8$\times$ higher read/write latencies, with 97.8\% link bandwidth being taken (even though both have the same amount of outstanding bytes). Third, the hardware prefetching implicitly induces many more loads into the target adapter under regular access patterns and predictable data locality, which is completely transparent to the CXL fabric. We observed that while the switch downstream port is not congested, the target adapter sees twice as much traffic, causing dramatic queue build-up.


Therefore, the switched CXL memory pooling lacks a transport layer that can effectively manage end-to-end fabric bandwidth among competing memory flows and enable application-specific traffic control.
However, realizing this is non-trivial. First, CXL packets (FLITs) traverse through the CPU pipeline, system bus, switching fabric, and remote DIMM, whose transmissions are implicit. Specifically, a load/store instruction is issued depending on data locality and the execution condition of micro-architectural components, and its response directly resumes the stalled processor execution without signaling. Second, CXL adapters and switches employ an ultra-fast data plane with limited in-network computing capabilities, whose hardware architecture is opaque, eluding any active traffic control mechanisms. Third, monitoring the data transmission performance requires us to bridge the gap between high-level memory streams and low-level hop-by-hop link-layer credits. The problem is further exacerbated by the inherent nature of the lossless network and the massive amount of application-induced memory requests.

We design and implement a new transport layer (dubbed {\bf \sys}) atop today's \texttt{CXL.mem} protocol stack, inspired by a seminal technique -- Core-Stateless Fair Queueing (CSFQ)~\cite{csfq-sigcomm98}. It achieves max-min bandwidth in the Internet for competing flows without maintaining per-flow states. We find that CSFQ is promising to tackle the above challenges in our context because (a) it is highly scalable, where core switches only maintain a few aggregated variables (e.g., arrival rate, accepted rate, and fair share rate) with fixed computing complexity;  (b) edge-driven, requiring minimal in-network support; (c) lightweight on the data plane,  whose traffic manipulation primitives (e.g., statistic bookkeeping, labeling, and packet dropping) are compute-efficient.

\sys comprises three pieces: (a) host programming interfaces that center around the \texttt{mchannel} system object with APIs to support unmodified applications; (b) the host runtime that runs the protocol stack, interacts with CXL fabric components, and performs rate control; and (c) in-fabric system extensions at the switch, adapter, and CXL DIMM, participating in the protocol execution. \sys probes the end-to-end bandwidth capability at runtime on the data plane, introduces new fabric congestion signals, and computes the per-\texttt{mchannel} transmission rate. {\it Key to \sys is a Sender-Driven Fabric-Informed transport protocol that admits just the right amount of CXL requests to each \texttt{mchannel} based on the estimated $Core \leftrightarrow DIMM_{CXL}$ bandwidth availability.} Specifically, we first apply the fluid model to the entire CXL fabric, figure out how to tailor CSFQ to our context, and derive the theoretical bound. Next, to handle the implicit transmission issue, \sys applies the time-based rate control that translates the end-to-end bandwidth usage and availability to the available running time. To avoid in-network data-plane operations, we push rate calculation to the edge, reserve a designated remote memory region for bookkeeping cross-host statistics, and translate in-network packet dropping to endhost admission control. We then follow the fluid model to determine the per-\texttt{mchannel} access speed and fair share rate and use a delay-based approach to probe the link bandwidth capacity. Further, we extend the \sys to support weight and multi-layer CXL switching.


We prototyped the \sys, ported several applications~\cite{workstealing-ppopp13, trie-accs07, gap-arxiv, mica-nsdi14} atop, and performed end-to-end evaluation over a real switched memory pooling setup. Our evaluations show that \sys fully uses the underlying link bandwidth, mitigates intra-host and cross-host performance interference, provides CSFQ-like fairness, adapts to the fabric bandwidth availability, and scales with the application demands. Applications running over \texttt{mchannel}s receive efficient multi-tenancy and achieve 2--3.5$\times$ improvements.

\section{Understanding Switched CXL MEM Pooling}
\label{sec:chara}


\subsection{Switched CXL Memory Pooling}
\label{subsec:chara-cxl-mempool}

\begin{figure}[tp]
    \includegraphics[width=\linewidth]{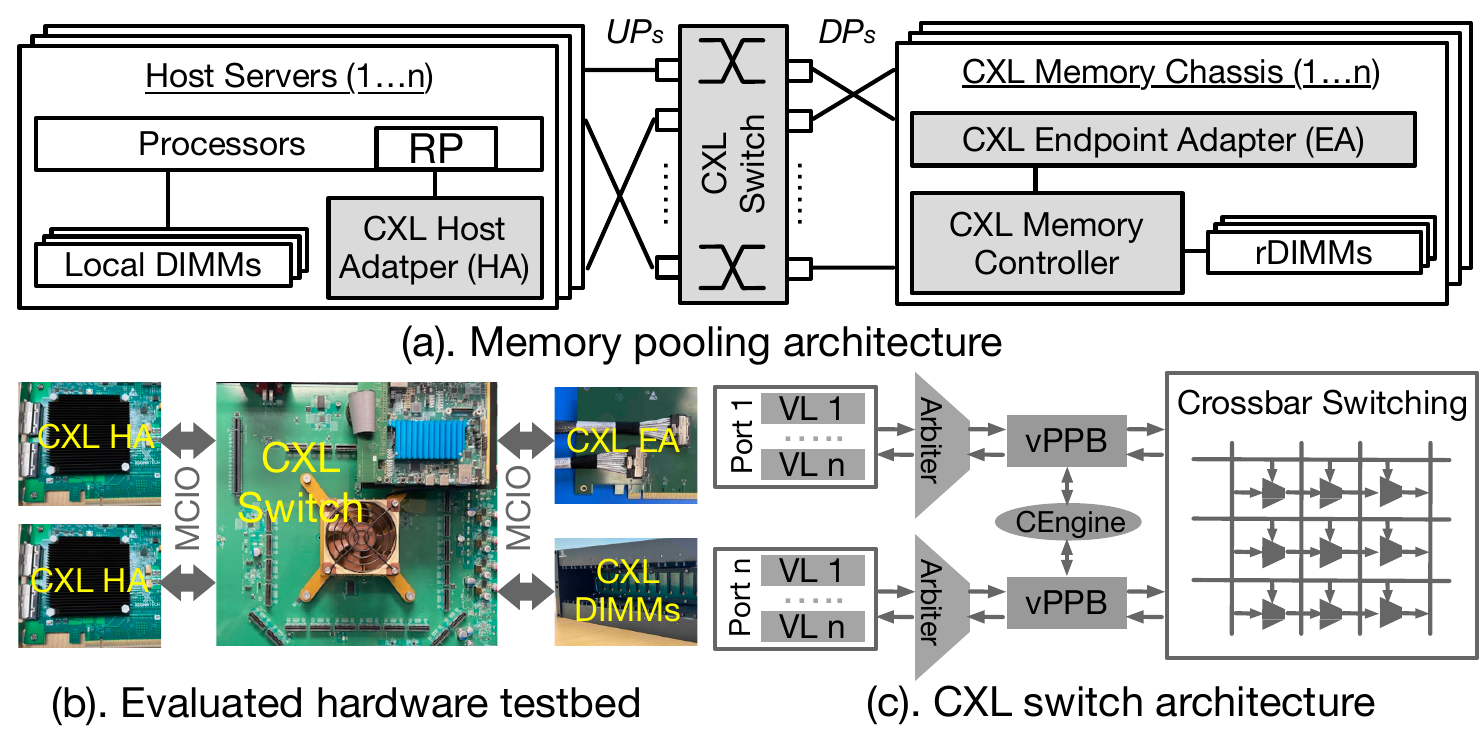}
    \vspace{-1.0\baselineskip}
    \caption{The architecture and hardware testbed of a switched CXL memory pool. (c) shows the architecture of a CXL switch. VL=Virtual Lane. vPPB=Virtual PCI-to-PCI Bridge.}
    \label{fig:hw-overall}
\end{figure}

CXL~\cite{cxl} is an emerging high-speed cluster interconnect built atop the physical layer of PCIe Express~\cite{pcie}. It provides a load/store interface for CPU, memory, and device communications. The interconnect uses the Flex Bus I/O architecture~\cite{cxl-spec}, and is organized into physical, data link, and transaction layers. CXL supports three types of channels: \texttt{CXL.cache}, \texttt{CXL.mem}, and  \texttt{CXL.io}.
Memory pooling is constructed using \texttt{CXL.mem} and the corresponding memory expanders (Type-3 device). Since CXL 2.0, CXL fabrics allow resource pooling via single-level and multi-tiered switching~\cite{cxl-spec}. Figure~\ref{fig:hw-overall} depicts the system architecture of a switched CXL memory pool.

\squishlist
    \item {\bf CXL Host Adapter.} It exposes access root ports (RPs) and cooperates with the host memory subsystem. The adapter converts load/store instructions into fabric-routable FLITs and transmits them over the wire. Upon responses, it parses the packets, obtains fetched read data or write completions, and delivers them to the processor pipeline.

    \item {\bf CXL Switch.} A CXL switch consists of upstream ports (UPs) for host adapters' connectivity, downstream ports (DPs) for remote memory, and internal forwarding tables for traffic orchestration. Upon initialization, it discovers all connected components, configures the routing structure, and fills table entries based on the topology. The switch carries CXL transactions on the data plane and employs some scheduling policies. CXL supports a hybrid of Port-Based Routing (PBR) and Hierarchy-Based Routing (HBR).

    \item {\bf CXL Endpoint Adapter.} It stays close to target memory devices, operating as a responder for remote memory. The adapter processes CXL protocols and converts between FLITs and memory commands. It also performs integrity checking, request steering (when multiple logic devices are used), and transmission speed synchronization.
    
    \item {\bf CXL Attached Memory.} The remote DIMMs (rDIMMs) are housed in a standalone chassis or appliance, including an SoC backplane, memory controllers, and a power supply. Each module uses a DDR-compatible PHY interface, supporting different kinds of memory media.
\squishend

\noindent
{\bf Evaluation Target.} Several memory appliances~\cite{samsung-cmmb, h3-falcon, unifabrix-max, omega-fabric, xconn-titan} have been developed, sampled, and tested in the past few years. We take the recent Titan platform~\cite{xconn-titan} from the XConn Tech as the evaluation target (Figure~\ref{fig:hw-overall}-b).
It consists of (1) ASIC-based host and endpoint adapters;
(2) an XConn B2 Apollo CXL switch with 14 upstream and downstream ports, 2 CEM~\cite{cem-pcie} slots, and 128 CXL2.0 lanes, supporting $\times$2, $\times$4, and $\times$8 bifurcations;
(3) a memory chassis using an MCIO/EDSFF backplane that holds up to 12 vendor-agnostic CXL DIMMs~\cite{smart-cxldimm, micron-cxldimm} under the compact EDSFF E1.S, E3.S, and E3.L form factors~\cite{jedec-cxldimm}; and (4) an MX8 board operating as the management host and fabric manager.


\noindent
{\bf Software Stack.} A switched CXL memory pooling platform has three software components: (1) a fabric manager running on a dedicated host, which interacts with each fabric hardware, discovers the system topology, enumerates hosts and target memory nodes, and monitors their living status; (2) the memory controller firmware for the remote DIMMs, which configures the communication ID, initializes its address space, and processes traversed data; (3) a device driver that makes remote memory as host-managed device memory (HDM) and exposes it as a CPUless NUMA node. The host OS fabricates an attached CXL DIMM as a memory node, manages it with object or tiered memory systems~\cite{directcxl-atc22, pond-asplos23, tpp-asplos23, cxlshm-sosp23, nomad-osdi24, aol-osdi25, collid-sosp24, hybridtier-asplos25}, and runs applications atop. After a CXL DIMM is mapped to the host memory subsystem, applications can access it via load/store instructions. Take the Intel x86 processor as an example. A memory read, missed from the last-level cache (LLC), fetches the corresponding cache line from the CXL memory. Memory writes hit the store buffer first, then are issued to the remote memory under eviction. Hardware/Software prefetch and cache coherence-induced events (like read-for-ownership) also cause data loads~\cite{melody-asplos25, pathfinder-sigcomm25}.

\subsection{CXL Switching Architecture}
\label{subsec:chara-cxl-switching}

A CXL switch provides high-performance and lossless connectivity between upstream and downstream ports (Figure~\ref{fig:hw-overall}-c). Data is transmitted at the FLIT granularity, a fixed-sized amount of data traversed over the underlying link, such as 68B and 256B. An incoming FLIT arrives at one virtual lane (VL) and is then forwarded to an arbiter (multiplexer). Next, the FLIT is delivered to a vPPB (virtual PCI-to-PCI bridge) module, acting as a logical connection point to accommodate different types of CXL devices and facilitate the routing. vPPBs can also be organized hierarchically to construct multiple VCS (virtual CXL Switch) within a physical switch. There is a credit engine (CEngine) interacting with all vPPBs and running (1) a hop-by-hop credit-based flow control~\cite{cfc-network95, cfc-sigcomm94}; (2) a credit update protocol (like N23) for reliable device-switch and switch-switch coordination~\cite{cfc-sigcomm94}; and (3) an adaptive and statistical credit allocation scheme to maximize bandwidth usage~\cite{cfc-network95}. Last, most of today's CXL switches (like XConn Apollo XC50256~\cite{xconn-titan} and Omega Fabric~\cite{omega-fabric}) employ a crossbar topology, where the routing logic is determined by the fabric manager during the memory pool initialization. Note that (1) the CXL switch is nearly bufferless but still incurs transmission stall when credit starvation happens; (2) there is little programmability on the switch data-plane, unlike Ethernet ones offering some in-network primitives in the traffic manager; (3) host and endpoint adapters usually adopt a similar switching architecture, just with much fewer ports. 

\subsection{Characterizing Switched CXL Memory Pooling}
\label{subsec:chara-issues}

\begin{figure}[t!]
    \centering
    \includegraphics[width=\linewidth]{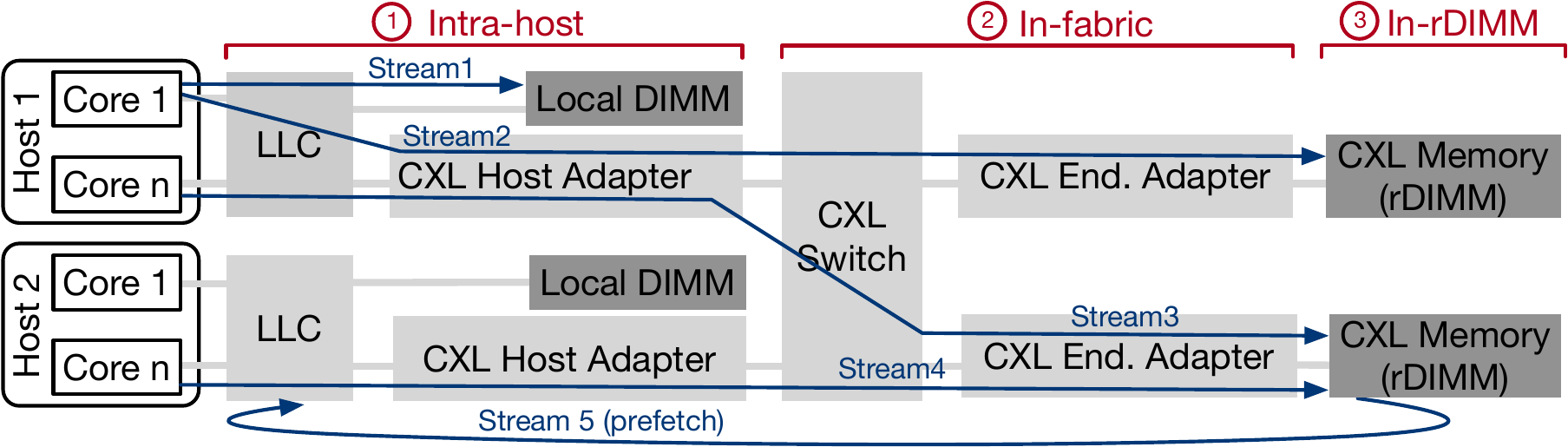}
    \vspace{-1.25\baselineskip}
    \caption{Different access contention points along the CXL data path under switched CXL memory pooling.}
    \vspace{-0.25\baselineskip}
    \label{fig:path-contention}
\end{figure}

Researchers have studied extensively about a direct-attached CXL Type-3 memory expander~\cite{cxl-chara-micro23, cxl-against-hotnets23, fcc-hotos23, directcxl-atc22, cxlintro-arxiv23, melody-asplos25, pathfinder-sigcomm25}. This section characterizes the performance of switched CXL memory pooling with a focus on analyzing the performance interference. Figure~\ref{fig:path-contention} presents how intra- and inter-host memory requests would interleave at different locations along the CXL data path in a switched memory pool. We configure experiments to locate the contention points, quantify how performance isolation is affected, and explore the root causes.


\noindent
{\bf Experimental Methodology.} We use 2U Intel Emerald Rapids servers as hosts, where each contains two Xeon Gold 6530 CPUs, 1536GB DDR5 memory, and two Samsung 9MA3 960GB NVMe drives. Each processor has 32 cores running at 2.1GHz and 160MB LLC. All the hosts run Ubuntu 24.04. Each server is equipped with one CXL host adapter enclosing two $\times$8 CXL ports. Using Intel MLC~\cite{intel-mlc}, we observe that the server achieves 220.5ns and 47.2GB/s when accessing the switched CXL memory pool. We developed a micro-benchmark (similar to~\cite{spa-asplos25, pchase, cxl-chara-micro23}) that can launch any number of memory streams between host cores and local/CXL DIMMs with different access patterns. It supports various configurations, such as enabling/disabling prefetching, issuing temporal read/write and non-temporal write, injecting NOP instructions, and performing random/sequential/strided accesses. The benchmark uses data and instruction synchronization barriers to complete pending reads and writes in the CPU pipeline before starting the test. When running, it spawns several pinned threads, initiates memory streams, issues reads/writes, and collects execution statistics.


\begin{figure*} [t!]
	\centering
	\subfloat[Uncore contention.]{
	\begin{minipage}{0.32\textwidth}
	\includegraphics[width=\linewidth]{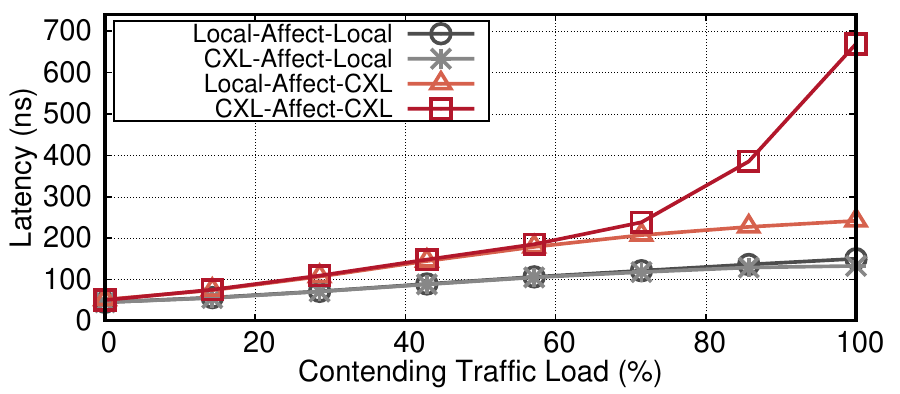}
        \end{minipage}\hfill
        }
	\subfloat[Host adapter contention (latency).]{
	\begin{minipage}{0.32\textwidth}
	\includegraphics[width=\linewidth]{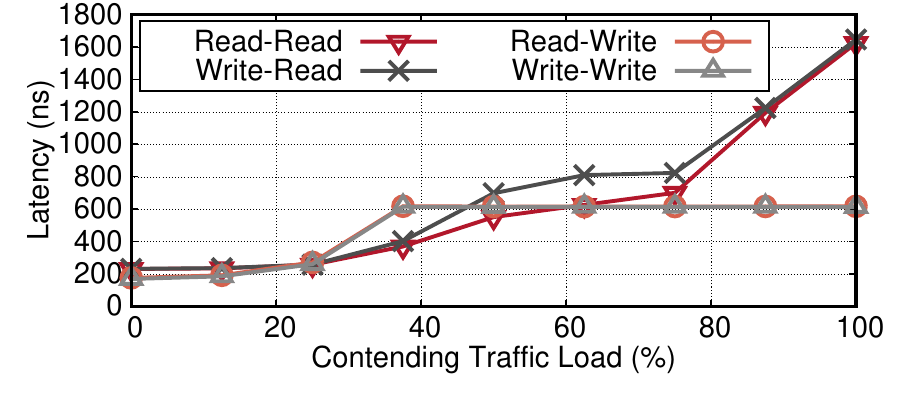}
        \end{minipage}\hfill
        }
        \subfloat[Host adapter contention (bandwidth).]{
	\begin{minipage}{0.32\textwidth}
	\includegraphics[width=\linewidth]{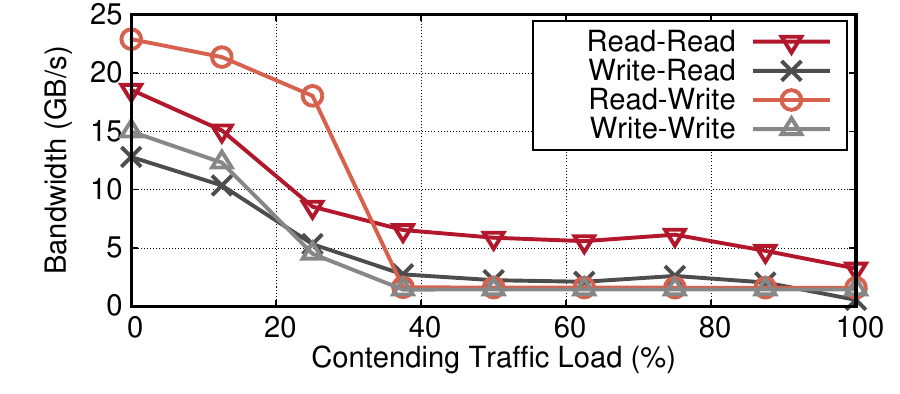}
        \end{minipage}\hfill
        }
        \vspace{-1.0\baselineskip}
        \caption{(a) runs background threads on one socket, issues sequential accesses, and maxes out its memory bandwidth. We report random access latency. The ``Local-Affect-Local'' and ``CXL-Affect-Local'' cases are added for comparisons. (b)/(c) present the host contention issue. X--Y shows that the victim stream (X) is impacted by co-located streams (Y). All experiments use one $\times$8 CXL port.}
        \vspace{-1.0\baselineskip}
        \label{fig:rc1-intrahost}
\end{figure*}

\begin{figure*} [t!]
	\centering
	\subfloat[Host-rDIMM interference.]{
	\begin{minipage}{0.32\textwidth}
	\includegraphics[width=\linewidth]{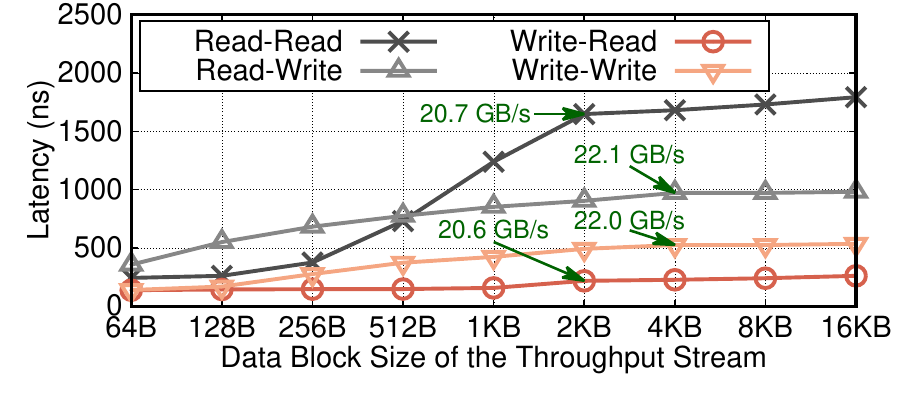}
        \end{minipage}\hfill
        }
        \subfloat[rDIMM-rDIMM interference.]{
	\begin{minipage}{0.32\textwidth}
	\includegraphics[width=\linewidth]{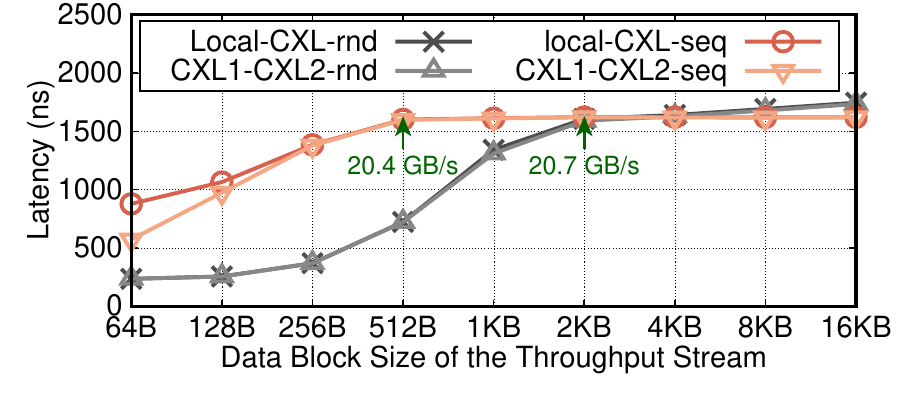}
        \end{minipage}\hfill
        }
        \subfloat[Bandwidth partition.]{
	\begin{minipage}{0.32\textwidth}
	\includegraphics[width=\linewidth]{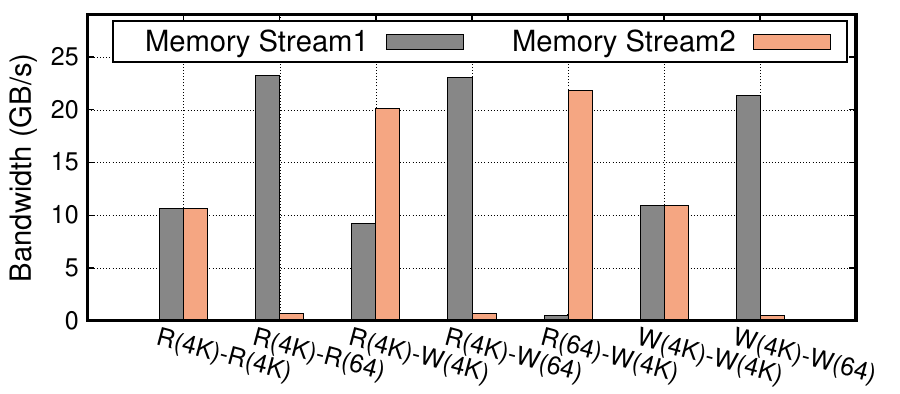}
        \end{minipage}\hfill
        }
        \vspace{-1.0\baselineskip}
        \caption{In (a), X--Y describes the case that a latency stream X is co-located with throughput streams (Y), where we report the average latency of X. (b) presents the latency results of four data movements. We add the local-rDIMM scenario for comparison. rnd=Random. seq=Sequential. In (c), we run two throughput streams each over four cores, issuing 64B/4KB read/write requests.}
        \vspace{-0.25\baselineskip}
        \label{fig:cxl-concur}
\end{figure*}

\noindent
{\bf Issue \#1: Intra-host Contention.} Host uncore and CXL host adapter are the first two contending locations. In an Intel X86 processor, the uncore--enclosing LLC, CHA (Caching and Home Agent), and FlexBus, connected via a mesh NoC~\cite{intel-sapphire-isscc22}--is shared by local and CXL memory requests, e.g., stream1 and stream2 in Figure~\ref{fig:path-contention}. We configure two kinds of threads: \emph{T1}, whose working set fits into the LLC, accessing local or CXL memory and measuring latency; and \emph{T2}, which generates competing traffic with gradually increased loads by adding the number of threads without overwhelming the host adapter's bandwidth capacity. As shown in Figure~\ref{fig:rc1-intrahost}-a, when the FlexBus is congested, accessing the CXL memory experiences up to 13.4$\times$ latency increases, rising from 50.1ns to 669.4ns. Using the recent PathFinder tool~\cite{pathfinder-sigcomm25}, we find out that this is because significant queueing happens at the M2PCIe ingress, which slows down CXL request transmission. When contentions occur at the LLC/CHA under mixed local and CXL traffic, we observe that both the LLC miss rate and the total amount of cache snooping requests greatly increase, yielding at most 2.9$\times$ CXL access latency increases.


Next, we analyze the host adapter contention by issuing cache-bypassed CXL requests (stream2 and stream3 in Figure~\ref{fig:path-contention}). We set up two memory streams: \emph{T1} accesses CXL memory with one outstanding load/store (i.e., victim memory stream); \emph{T2} reads/writes to remote memory, with the number of outstanding accesses gradually increased.
As shown in Figures~\ref{fig:rc1-intrahost}-b/c, T1 starts to experience a performance drop even when the total traffic only reaches 30\%, far below the adapter's bandwidth capacity. By discussing with the device vendor, we learn that this is because the host adapter schedules CXL FLITs over several VLs of the virtual lane in a best-effort fashion, completely agnostic of how requests are issued from host cores, which causes skewed cross-lane FLIT distribution.
When the adapter is nearly saturated, we observe that in the Read(T1)-Read(T2) case, the victim stream's bandwidth is reduced by 82.6\%, with latencies increased from 230.2ns to 1624.2ns. The other three cases are similar.


\noindent
{\bf Issue \#2: In-fabric Congestion.} The CXL switch is the second interference point (stream3 and stream4 in Figure~\ref{fig:path-contention}). Since the CXL switching relies on a hop-by-hop credit-based flow control~\cite{cfc-sigcomm94, cfc-network95, rpcibench-nsdi24}, the amount of credits that a downstream entity receives is proportional to its bandwidth usage and the contention degree at the upstream device. Hence, when a small-sized request stream competes with a large one that causes credit over-subscription, it would experience stalls due to delayed credit replenishment. To quantify this, we configure two kinds of random access memory streams: the latency stream that issues one memory read/write at a time, and the throughput one, which generates larger data blocks, yielding multiple outstanding memory transactions. 

We intermix small and large memory requests and deploy memory streams in two directions: host$\rightarrow$rDIMM and rDIMM$\rightarrow$rDIMM. Regarding the host$\rightarrow$rDIMM case, as shown in Figure~\ref{fig:cxl-concur}-a, when the fabric is under-utilized, a 64B read mixed with 128B reads and writes causes a 7.5\% and 54.2\% latency increase, respectively. In the case of a 64B write contending with throughput streams, when the data block size is 256B, its latency is increased by 6.2\% and 98.6\%. This is due to the head-of-line (HoL) blocking effect at the upstream port. When bandwidth over-subscription happens, one would experience considerable performance drops. For example, when interleaved with a 4KB memory request, a 64B read/write experiences a 6.9$\times$/3.8$\times$ slowdown. The rDIMM$\rightarrow$rDIMM scenario presents similar behaviors (Figure~\ref{fig:cxl-concur}-b), where the 64B CXL memory access latency starts to rise when the data block size is above 256B. 

Next, we deploy two competing throughput streams, vary their request configurations, and explore how bandwidth is allocated in different cases. As shown in Figure~\ref{fig:cxl-concur}-c, we find that the bandwidth a memory stream receives is mainly proportional to its data block size, regardless of request type. For example, when two 4KB read(write) streams interleave, each would achieve 10.6(11.0) GB/s. However, if a 4KB stream is interleaved with a 64B one, it can take nearly 97.8\% of the total bandwidth. These results make sense for two reasons. First, memory requests from one memory stream are synchronously served one by one. The number of outstanding cacheline-sized reads/writes depends on the data block size. Second, when multiple concurrent streams compete for the fabric, the CXL switch/adapter mainly performs round-robin scheduling to decide the next issuing request. Thus, the stream with more pending transactions has more opportunities to be scheduled, yielding higher bandwidth.

\noindent
{\bf Issue \#3: Unmanaged Host-rDIMM Interaction.}
The endpoint adapter 
provisions enough bandwidth for CXL DIMMs. Since it connects directly to the switch, access congestion is first resolved at the downstream port, which should leave the adapter and rDIMM free from contending traffic. However, we find out this is sometimes not the case. Even though we control the aggregated traffic across all hosts to be lower than the link/port bandwidth, contention still happens. This is because host prefetching, under regular access patterns,
implicitly generates more memory requests, which are transparent to the CXL fabric but cause bandwidth over-subscription.


\begin{table}[t!]
    \scriptsize
    \begin{tabular}{C{12mm}|| C{11mm} | C{14mm} | C{11mm} | C{14mm}}
    \hline
    \textbf{Stride Len.} & \textbf{RD Lat.} & \textbf{RD Th.} & \textbf{WR Lat.} & \textbf{WR Th.} \\
    \hline
    \textbf{1} & 10.9 ns & 21.9 GB/s & 10.8 ns & 22.0 GB/s \\
    \hline
    \textbf{2} & 15.6 ns & 15.3 GB/s & 13.2 ns & 18.0 GB/s \\
    \hline
    \textbf{4} & 18.7 ns & 12.7 GB/s & 13.3 ns & 17.9 GB/s \\
    \hline
    \textbf{6} & 18.2 ns & 13.1 GB/s & 12.3 ns & 19.4 GB/s \\
    \hline
    \textbf{8} & 18.6 ns & 12.8 GB/s & 12.4 ns & 19.2 GB/s \\
    \hline
    \end{tabular}
    \vspace{-1.0\baselineskip}
    \caption{CXL memory performance under a fixed stride size $X$. The distance between two consecutive requests is $X \times 64$ bytes.}
    \label{tbl:strided-perf}
    \vspace{-0.5\baselineskip}
\end{table}

\begin{figure} [t!]
	\centering
        \subfloat[Latency.]{
	\begin{minipage}{0.24\textwidth}
	\includegraphics[width=\linewidth]{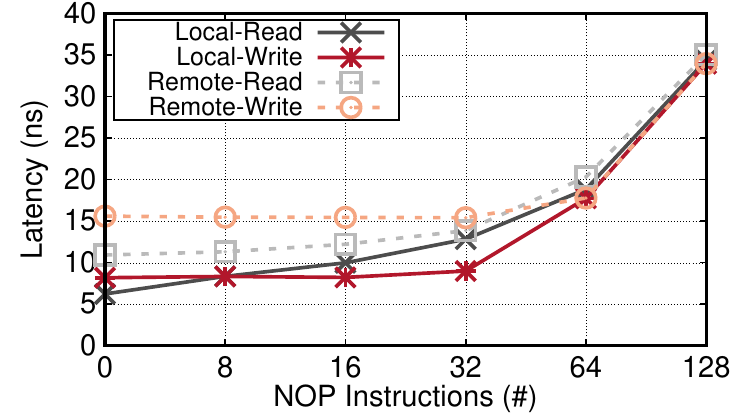}
        \end{minipage}\hfill
        }
        \subfloat[Bandwidth.]{
	\begin{minipage}{0.24\textwidth}
	\includegraphics[width=\linewidth]{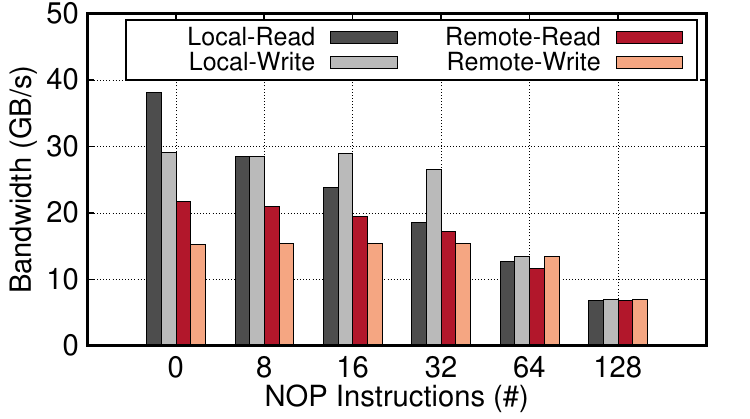}
        \end{minipage}\hfill
        }
        \vspace{-1.0\baselineskip}
        \caption{Performance when perturbing instruction prefetching. }
        \vspace{-0.25\baselineskip}
        \label{fig:cxl-prefetch}
\end{figure}

We configure our microbenchmark to issue strided memory reads and writes. Table~\ref{tbl:strided-perf} presents the latency and throughput as the stride length increases from 1 to 8. When the stride length is 1, compared with the random access, remote memory read and write achieve 10.9ns and 10.8ns (hitting in the L1 and L2 cache), sustaining at 21.9GB/s and 22.0GB/s bandwidth. Apparently, the access locality results in multiple cacheline reads and writes issuing concurrently to hide latency. When the stride length increases to 8, the read and write bandwidth drops to 12.8 GB/s and 19.2 GB/s. Further, we perturb the CPU prefetching effect by issuing NOP instructions between two consecutive memory accesses and measure its performance impact. We find that both local and CXL memory performance drops significantly when inserting more NOP instructions (Figure~\ref{fig:cxl-prefetch}). When prefetching helps little, a remote memory read/write matches the local one. Therefore, host prefetching causes unmanaged host-rDIMM interaction, which would introduce much more traffic than expected and cause contention within the target memory pool.

\subsection{Problem and Challenges}
\label{subsec:chara-problem-challenges}

Our characterization study unearths and quantifies three issues (i.e., intra-host contention, in-fabric congestion, and host-rDIMM interaction) that cause performance interference in a switched CXL memory pool. {\it The fundamental problem is that the data path between a host core and a remote CXL DIMM is shared among concurrent memory requests without explicit performance control.} As such, any on-path entities, like the host memory subsystem, switch, and host/endpoint adapter, could become a communication choke point, which would break the target latency and bandwidth of a memory stream. This calls for a new transport layer for switched memory pooling that effectively allocates communication resources based on the application requirements. Realizing this incurs the following three challenges that have not been tackled before.

\squishlist

    \item {\bf \#1: Implicit transmission.} Under switched memory pooling, a CXL request traverses across the CPU pipeline, cache hierarchy, system bus, adapters, switching fabric, and remote DIMM. Whether and when to issue a CXL load/store transaction is affected by several factors, such as data locality (L1D, L2, and LLC), queueing occupancy at different micro-architectural components (such as store buffer and line fill buffer), and hardware prefetching. Similarly, a data response returns to the host memory subsystem and directly resumes the processor execution. Therefore, tracking CXL requests, measuring the per-flow performance, and controlling the transmission rate become non-trivial.
    
    \item {\bf \#2: Hardware non-programmability and opaqueness.} To sustain at sub-microsecond latency and hundreds of Gigabytes per second of bandwidth (CXL 3.2~\cite{cxl-spec}), CXL adapters and switches streamline their data plane and offer nearly no reconfigurability and telemetry capability, making our conventional in-network transport design~\cite{abm-sigcomm22, dcqcn-sigcomm15, hpcc-sigcomm19, aeolus-sigcomm20, harmony-nsdi24} impractical. Further, there is no consensus system model or open implementation standard (like PISA~\cite{rmt-sigcomm13}).

    \item {\bf \#3: Fine-grained cross-layer traffic monitoring.} Measuring end-to-end bandwidth availability is already challenging enough since this requires tracking how many link-layer credits are allocated across the entire data path. However, in a switched pool, the transport layer has to further break down this information at a finer granularity to individual host cores/threads. As described above, the round-trip delay is impractical to obtain, given the integrated pipeline. The inherent nature of lossless fabric (back-pressure and credit starvation) further exacerbates the issue.

\squishend


\section{\sys: a Designated Memory Lane}
\label{sec:mchannel}

This section introduces the \sys transport, including semantics, APIs, and system model. Our system design goals:


\squishlist

    \item {\bf High Utilization.} \sys should fully use the bandwidth at any vantage point and link. It should be able to ramp up the CXL memory access speed for other needed applications when some bandwidth becomes available.

    \item {\bf Efficient Multi-tenancy.} \sys should mitigate the performance interference across the end-to-end CXL data path ($\S$\ref{subsec:chara-issues}). It should achieve low (tail) latency and approximate Max-Min fairness under contention.
        
    \item {\bf Low overheads.} \sys should have little impact on the de facto CXL load and store access. It should incur few memory footprints and consume tolerable host CPU cycles when holding existing applications.
    
\squishend

\subsection{Overview}
\label{subsec:mchannel-overview}

\sys is a transport layer that orchestrates remote memory accesses between host cores and CXL DIMMs in a switched CXL memory pool. It offers a performance-controlled data pipe (\texttt{mchannel}), provisions proper communication resources based on workload demands and remote DIMM's free bandwidth, and mitigates interference from contending memory streams. A \texttt{mchannel}, established between a host core ($C_i$) and a CXL DIMM ($rDIMM_j$), is associated with a dedicated application process. One can also group multiple \texttt{mchannels} for each application (discussed in $\S$\ref{subsec:proto-ext}).

Key to \sys is a {\it Sender-Driven Fabric-Informed} transport protocol that admits just enough CXL requests to the switched memory pool based on the estimated $C_i \leftrightarrow rDIMM_j$ bandwidth availability.
Essentially, it probes the end-to-end bandwidth capability at runtime on the data plane, introduces new fabric congestion signals, computes the per-\texttt{mchannel} transmission rate
and admits an adequate amount of CXL load/store commands to the memory pool. \sys encompasses three major pieces. One is the programming interface that allows porting unmodified applications.
The second is the host runtime that runs the protocol stack, interacts with other CXL fabric components, and performs rate control. The third is in-fabric system extensions at the switch, adapter, and CXL DIMM, which participate in the protocol processing.

\subsection{The \texttt{mchannel} Semantics}
\label{subsec:mchannel-semantics}

\sys exposes the \texttt{mchannel} as a system object from the host OS perspective, consisting of the following attributes:

\squishlist
    \item {\it Host/Core ID and Registered Memory Node}, specifying the requester (source) and responder (destination) of a \sys. Akin to commodity systems~\cite{omega-fabric, unifabrix-max, samsung-cmmb}, we support DAX memory, mounted as a CPUless NUMA node.

    \item {\it CXL Data Path}, describing the end-to-end data path between $C_i$ and $rDIMM_j$, provided by the fabric manager, including host adapter, switches, and endpoint adapter.
    
    \item {\it Channel Representation}, like \texttt{struct sock}, instantiated by our host runtime, serving as an intermediate connection point between applications and the transport protocol.  
    
    \item {\it Performance Attributes}, like bandwidth envelope and interference degree. Developers specify them upon registration or at runtime. When omitted, the \texttt{mchannel} is labeled as best-effort and uses what bandwidth is available.

\squishend

\begin{figure}[tp]
    \centering
    \includegraphics[width=\linewidth]{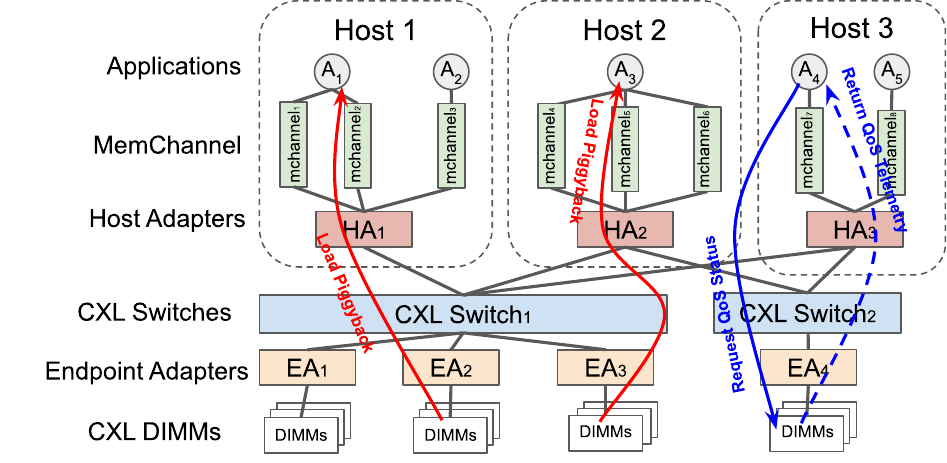}
    \vspace{-1.5\baselineskip}
    \caption{The system model of \sys.}
    \label{fig:sys-model}
    \vspace{-0.25\baselineskip}    
\end{figure}

\subsection{The \texttt{mchannel} Interface}
\label{subsec:mchannel-intf}

\sys provides a user-space CLI (Command-Line Interface), similar to \texttt{numactl}, which onboards the unmodified applications and controls their execution behaviors to achieve the performance target. \sys runtime offers four lightweight APIs to control the remote memory access:


\squishlist

\item \texttt{mchannel\_init} creates \texttt{mchannels} and takes user-specified configurations, such as which applications to run, host core mappings, remote memory nodes, and performance attributes. It then identifies the CXL data path for each \texttt{mchannel}, allocates local memory for \texttt{mchannel} metadata, maps shared CXL memory for inter-host coordination, and launches the applications as subprocesses.


\item \texttt{mchannel\_open} opens a \texttt{mchannel} to a rDIMM in the pool, invoked when a new application thread is launched. It allocates the per-thread \texttt{mchannel} states and resources, and then arms a POSIX timer with a scheduling window $T_W$. When the timer expires, it invokes \texttt{mchannel\_sched} as the signal handler and readjusts the bandwidth allocation.
It also uses the \texttt{m\_bind} system call to ensure the application places its data on the remote DIMM specified by the \texttt{mchannel}, sets the thread’s core affinity based on the host core mappings, and opens as well as maps the performance counters for CXL bandwidth monitoring.

\item \texttt{mchannel\_close} closes the \texttt{mchannel} to the remote DIMM, cleans up allocated resources, and unsets the timer signal handler.  It is called when the application thread exits.

\item \texttt{mchannel\_sched} enforces application-specific bandwidth allocation via our transport protocol ($\S$\ref{sec:proto}), which monitors the remote pool access bandwidth and then controls application execution to meet the application performance needs.
For bandwidth measurement, we use off-core response (OCR) architectural event counters to track three types of CXL requests~\cite{pathfinder-sigcomm25, melody-asplos25}: demand reads, reads for ownership, and hardware prefetches. The results are collected via \texttt{rdpmc} commands and aggregated as the remote bandwidth. For bandwidth control, during a scheduling window $T_W$, we adjust the POSIX timer expiration such that the application thread runs for $T_R$ and sleeps for $T_W - T_R$.


\squishend

\sys relies on code injection techniques~\cite{ci-wiki}, such as \texttt{LD\_PRELOAD} and \texttt{ptrace}, to support unmodified applications. Specifically, \sys injects a dynamic link library (DLL) into applications when launching them. Inside the DLL constructor and \texttt{pthread\_create} interposed by DLL, \sys calls \texttt{mchannel\_open}.
The \texttt{mchannel} interfaces incur low overheads,
because \texttt{mchannel\_sched} is invoked only once per scheduling window $T_W$ and halts application execution only once during that window. In addition, we utilize lockless data structures to synchronize between \texttt{mchannels} and the \texttt{rdpmc} command to read performance counters inside \texttt{mchannel\_sched} instead of system calls.

\subsection{System Model}
\label{subsec:channel-model}

We divide our running environment into six layers (Figure~\ref{fig:sys-model}): applications, \textbf{mchannel}s, host adapters, CXL switches, endpoint adapters, and CXL DIMMs. Our \sys layer carries application-induced CXL memory requests 
and delivers them to the host adapter. The CXL FLIT traverses the underlying fabric as it is. Following the recent CXL specification~\cite{cxl-spec}, the memory device continuously reports service loads, reflecting its internal queueing status. There are two ways: (a) {\it piggyback in memory responses} (red line in Figure~\ref{fig:sys-model}), where the memory transaction response reserves 2 bits to indicate four levels of loads: light load, optimal load, moderate overload, and server overload; (b) {\it QoS telemetry messages} (blue line in Figure~\ref{fig:sys-model}), i.e., the host issues explicit QoS status queries through the CXL Component Command Interface (CCI), whereas the memory expander (rDIMM) returns the average load of the last profiling epoch.

\section{An CSFQ-Inspired Transport}
\label{sec:proto}

This section describes our core transport protocol 
and shows how we address the above challenges (\(\S\)\ref{subsec:chara-problem-challenges}).

\subsection{Why Core-Stateless Fair Queueing}
\label{subsec:proto-csfq}

Core-Stateless Fair Queueing (CSFQ)~\cite{csfq-sigcomm98}, a seminal work in computer networks, introduced a fair queueing algorithm for the Internet several decades ago. 
It achieves max-min fair bandwidth allocation among competing flows without maintaining per-flow states. The algorithm 
divides the rate control logic between the networking edge and core, where 
(1) edges 
calculate the flow arrival rate and label the rates into packets; 
(2) the core 
estimates the fair rate iteratively and probabilistically drops packets to achieve the fair share rate. 
CSFQ is (a) 
scalable, where core routers only maintain a few aggregated variables (e.g., arrival rate, accepted rate, and fair share rate) with fixed computing complexity; (b) edge-driven, requiring minimal in-network support; (c) lightweight on the data plane, the traffic manipulation primitives (e.g., statistic bookkeeping, labeling, and packet dropping) are compute-efficient, making it promising to address our problem.

However, naively applying CSFQ and its successor HCSFQ~\cite{hcsfq-nsdi21} is still not feasible for three reasons. First, these approaches take packet drops as a congestion signal, but CXL fabric is a lossless network. Second, core switches require the packet-dropping primitive for traffic regulation, which is not supported on CXL switches. Third, CSFQ needs accurate flow rate and link bandwidth availability estimations, but CXL switching lacks such capabilities. Therefore, akin to CSFQ and HCSFQ, we first apply the fluid model to the CXL switching, derive the theoretical guarantees when achieving the max-min fairness (\(\S\)\ref{subsec:proto-fluid}), and redesign the transport algorithm (\(\S\)\ref{subsec:proto-alg}). To handle the implicit transmission issue, \sys applies the time-based rate control that translates the end-to-end bandwidth usage and availability to the available running time. To avoid in-network data-plane operations, we push rate calculation to the edge, reserve a designated remote memory region for bookkeeping cross-host statistics, and translate in-network packet dropping to endhost admission control. We then follow the fluid model to determine the per-\texttt{mchannel} access speed and fair share rate and use a delay-based approach to probe the link bandwidth capacity.

\setlength{\abovedisplayskip}{6pt plus 2pt minus 2pt} 
\setlength{\belowdisplayskip}{6pt plus 2pt minus 2pt} 
\setlength{\abovedisplayshortskip}{6pt plus 2pt minus 2pt} 
\setlength{\belowdisplayshortskip}{6pt plus 2pt minus 2pt} 

\subsection{Applying Fluid Model to CXL Switching}
\label{subsec:proto-fluid}

We apply the fluid model to formalize switched CXL memory pooling systems. The system is viewed as a directed acyclic graph \(G(V, E)\), where \(V\) represents system components (e.g., the ones in Figure~\ref{fig:sys-model}),
and \(E\) represents the traffic flows between components. We assume the flows are continuous memory streams from a host core to a remote CXL DIMM. We consider a single layer of switching in the following discussion, but our analysis applies to multiple layers.
A flow starts from a memory channel, goes through a bridge and a switch, and reaches the CXL DIMM. 
Therefore, graph \(G(V, E)\) is a complete multipartite graph, where the vertex set is partitioned into four disjoint subsets \(\{V_C, V_H, V_S, V_E\}\) for channels, host adapters, switches, and endpoint adapters, respectively.


Each CXL hardware component has a maximum link transmission capacity \(C_v\). 
Let \(\alpha_v\) be the fair share rate a node \(v\) allocated to its successors. We know the application access demand \(D_c\) for channel \(v_c \in V_C\) via our mechanism (\(\S\)~\ref{subsec:proto-alg}).
For the edge between a channel and a host adapter, the aggregated demand is the channel's demand, e.g., \(D_{ch}=D_c; v_h \in V_H \). The demand for other edges can be calculated recursively.
\begin{equation}
\label{eq:demand} 
D_{vw}= \sum_{e_{uv}\in p;  \forall p \in P_{vw}} \min(\alpha_v, D_{uv})
\end{equation}
\(p = (e_{ch}, e_{hs}, e_{se}); e_{ch}, e_{hs}, e_{se} \in E\) describes a flow path. \(P_{vw} = \{p|e_{vw} \in p\}\) is all the flows going through edge \(e_{vw}\). Therefore, under contention, where the aggregated memory demand \(D_v = \sum_{e_{uv} \in E} D_{uv} > C_v\), one can achieve the max-min fair bandwidth allocation among competing flows when \(\alpha_v\) is the unique solution to the following equation.
\begin{equation}
\label{eq:alpha-1}
C_v = \sum_{e_{uv}\in E} min(\alpha_v, D_{uv})
\end{equation}
Without contention (e.g., \(D_v \leq C_v\)), all demands can be satisfied by setting \(\alpha_v\) to the max, i.e., \(\alpha_v = \max_{e_{uv}\in E} D_{uv}\).
There are two key differences when applying the fluid model in our case compared with CSFQ. First, memory access rates remain the same along the flow path. As a result, we compute memory access demand instead of using flow arrival rates for edges and vertices. Second, we use the link transmission capacity at each node rather than the default maximum bandwidth specified in the hardware documentation. This is because the number of FLITs a CXL port of an adapter or switch can handle and transmit is not static but depends on the queuing conditions and internal execution status.


\subsection{The Transport Algorithm}
\label{subsec:proto-alg}

\begin{algorithm}[t]
\caption{Transport Algorithm.}\label{alg:proto-alg}
\small
\begin{algorithmic}[1]    

    \Procedure{mchannel_sched}{mchannel $c$}
        \State \(n\) = read_perf_counters() \Comment{CXL memory access count}
        \State \(c.demand\) = estimate_rate(\(c.demand\), \(n\), \(c.T_W\));
        \Comment{Eq.~\ref{eq:rate-est}}
        \State \(c.rate\) = estimate_rate(\(c.rate\), \(n\), \(c.T_R\));  \Comment{Eq.~\ref{eq:rate-est}}
        \For {On_Path_Device \ \(dev\) \ in c.path}
            \State \(\alpha\) = estimate_\(\alpha\)(\(dev\));
            \State \(c.\alpha\) = min(\(\alpha\), \(c.\alpha\)); 
        \EndFor
    
        \State \(c.T_R\) = estimate_\(T_R\)(\(c.demand\), \(c.\alpha\), \( T_W\)); 
        \Comment{ Eq.~\ref{eq:time-alloc}}
        \State sleep(\(c.T_W - c.T_R\))  \Comment{Yield to other mchannels}
    \EndProcedure

    \Procedure{estimate_\(\alpha\)}{On_Path_Device $dev$}
        \If {cur_time \(\ge\) start_time + \(K\)} 
            \State \(F\), \(D\) = 0, 0; \Comment{local rate \(F\), local demand \(D\)}
            \For {mchannel \ \(c\) \ in \(dev.mchannels\)}
                \State \(F\) += \(c.rate\);
                \State \(D\) = max(\(D\), \(c.demand\));
            \EndFor
            \State \(dev.F\), \(dev.D\) = agg_rate(\(F\), \(D\)); \Comment{via shared memory}
            \If {\(localhost\) is \(dev.host\)}
            \State \(dev.\alpha\) = min(\(dev.\alpha \times \frac{dev.C}{dev.F}\), \(dev.D\));  \Comment {Eq.~\ref{eq:alpha-est}}
            \EndIf
            \State start_time = cur_time;
        \EndIf         
        \State \Return \(dev.\alpha\);        
    \EndProcedure  

    \Procedure{adjust_capacity}{On_Path_Device $dev$}
    \If {\(dev.F < dev.C\) and \(dev.l > \delta\)}
        \State \(dev.C\) = \(dev.C \times (1 - \frac{dev.l-\delta}{2(1-\delta)})\) \Comment{Multiplicative Decrease}
    \ElsIf {\(dev.l \le \delta\)}
        \State \(dev.C\) = \(dev.C + \lambda \times (\delta -dev.l)\times K \) \Comment{Additive Increase}
    \EndIf
    \State \(dev.C\) = max(\(dev.C\), \(dev.\)max_capacity)
    \EndProcedure
\end{algorithmic}
\end{algorithm}

\noindent
\noindent
{\bf Access Rate and Demand Estimation.} 
We measure the number of remote cacheline accesses \(n\) during a scheduling window \(T_W\) by reading performance counters (§\ref{subsec:mchannel-intf}). 
The measured CXL memory access rate for \textit{mchannel} \(c\) is \(r_c' = \frac{nS}{T_W}\), where \(S\) denotes the cacheline size. \(r_c'\) reflects the amount of CXL bandwidth consumed by \textit{mchannel} \(c\) over \(T_W\). Similarly, the measured memory demand is \(D_c' = \frac{nS}{T_R}\), where \(T_R\) is the application’s running time within \(T_W\). \(D_c'\) indicates the memory bandwidth that \textit{mchannel} \(c\) would generate if it were not throttled by the transport algorithm. Measuring application demand in this manner is possible because the application runs at full speed without interruptions or slowdowns during \(T_R\), as our algorithm ensures that CXL components are not oversubscribed. Following the definition of CSFQ, we then estimate the memory access rate as the following equation, where \(K\) is a constant for adjusting the sampling window.
\begin{equation}
\label{eq:rate-est}
    r_c^{new} = (1-e^{-T_W/K}) r_c' + e^{-T_W/K} r_c^{old}
\end{equation}
We estimate the access demand \(D_c\) by substituting \(r_c'\) with \(D_c'\). Hence, hosts update \(r_c\) and \(D_c\) of each channel at the beginning of the scheduling function (ALG\ref{alg:proto-alg} L2–4). Since CXL fabric is lossless, the aggregated memory access rate can be calculated directly, rather than using Eq.~\ref{eq:rate-est} as in CSFQ.

\noindent
{\bf Fair Share Rate Estimation.} \sys then calculates the fair share rate \(\alpha\) on the host and issues the right amount of load/store instructions to avoid congestion. However, using Eq.~\ref{eq:alpha-1} to compute \(\alpha\) directly requires a host to know the aggregated memory demands on edges (\(D_{uv}\)) across all on-path hardware it traverses. Unlike Ethernet, where the arrival rate (i.e., demand) can be directly measured at the switch port, here it must be obtained through recursive calculations (Eq.~\ref{eq:demand}). Specifically, the memory demand of a CXL endpoint adapter depends on incoming memory streams from all corresponding upstream CXL switch ports, which in turn depend on incoming streams from upstream host adapters. This recursive dependency can easily lead to an accumulation of error margins. Furthermore, both demands and fair share rates must be shared across hosts, making the algorithm difficult to scale. Instead, we estimate the fair share rate based on the aggregated remote memory access rate \(F_v\) on the host for hardware node \(v\), defined as the number of load/store commands transmitted through \(v\). This approach is straightforward to compute and can be calibrated using hardware bandwidth telemetry.

We next describe how \sys performs fair-share rate estimation (Algorithm\ref{alg:proto-alg} L10--20). We define the aggregated memory access rate as \(F_v = \sum_{e_{uv} \in E} f_{uv}\), where \(f_{uv}\) is the flow on edge \(e_{uv}\) and \(f_{cb} = r_c\) for \(\forall v_c \in V_C, \forall v_h \in V_H\). Since the access rate is the same within a flow \(p\), the aggregated rate \(F_v\) is expressed as \(\sum_{e_{cb} \in p, \forall p \in P_v} r_c\), where \(P_v\) denotes all flows passing through node \(v\). Under congestion (Eq.~\ref{eq:alpha-1}), we estimate \(\alpha\) iteratively using the current congestion condition \(\frac{C_v}{F_v}\). If the estimated fair share rate exceeds the largest local demand \(\max_{c \in V_L} D_c\), there is no congestion within the local host, and thus \(\alpha\) is set to the maximum demand, i.e., \(\max_{e_{uv}\in E} D_{uv}\). 
\begin{equation}
\label{eq:alpha-est}
\alpha_{new} = \min( \alpha_{old} \frac{C_v}{F_v}, \max_{e_{cb} \in p, \forall p \in P_v} D_c )
\end{equation}
At the end of the sampling window \(K\), hosts calculate the fair share rates of all hardware components used by their \texttt{mchannels}. The aggregated memory access rate \(F_v\) is the only variable shared across hosts to compute the fair share rates for shared CXL hardware (e.g., switches and expanders). To disseminate this information, each host writes the local sum of the memory access rates for each hardware component to a small, dedicated shared memory address. The aggregated rate \(F_v\) can then be obtained by summing these local rates. The \texttt{mchannel} fair share rate \(\alpha_c\) is the minimum value along path \(p\) (i.e., \(\alpha_c = \min_{e_{uv} \in p} \alpha_v\)), which is then translated into the running time to control application execution.

 
\noindent
{\bf Time-based Rate Control.} 
\sys controls the FLIT transmission rate by adjusting the application thread running time \(T_R\) within each scheduling window \(T_W\) ($\S$\ref{subsec:mchannel-intf}). After running for \(T_R\), the application thread yields to another and remains in the waiting queue until the next scheduling window begins. A short scheduling window \(T_W\) enables fine-grained control over application execution but incurs high context-switching overhead due to frequent timer-expiration interrupts. Conversely, a larger \(T_W\) reduces scheduling overhead but causes tail latency increases
because requests are less tightly coordinated. We choose \(T_W = 100\,\mu\text{s}\) in our prototype, which strikes a balance between these two trade-offs.

Similar to TCP, a flow cannot transmit bytes when there is no free space in the congestion window. By temporarily stalling application execution, we can regulate the memory access rate to prevent congestion. To avoid temporal contention, application threads and their \texttt{mchannels} run asynchronously, even though they may have the same scheduling window. \sys estimates the CXL memory access size \(B\) during one scheduling window as \(D_c T_R\). To ensure fair sharing of CXL memory, the max-min fair remote memory size \(B_{fair}\) that an application can access during \(T_W\) is \(\min(\alpha_c, D_c) \times T_W\). The goal of our transport is to achieve max-min fair resource allocation without wasting CXL memory bandwidth. Therefore, we set \(B = B_{fair}\). Rearranging gives a fair running time during one scheduling window:
\begin{equation}
\label{eq:time-alloc}
T_R = \frac{\min(\alpha_c, D_c)T_W}{D_c}
\end{equation}

\noindent
{\bf New Congestion Signals.} In Ethernet, congestion can be readily inferred from explicit signals like packet loss.
However, such signals are largely unavailable in the CXL fabric. Instead, the CXL 3.2 specification~\cite{cxl-spec} mandates that memory expanders include a 2‑bit device‑internal load indication in memory responses (S2M), which encodes four levels of congestion typically derived from the expander’s internal queue occupancy. In addition, the specification strongly recommends that memory devices expose a QoS telemetry mechanism that periodically reports (1) egress‑port backpressure, indicating that one or more upstream queues between the expander and the host are saturated, and (2) temporary throughput reductions 
during DRAM refresh operations.

Congestion at the adapter can be queried at the host since the device is attached locally. For CXL DIMMs, we leverage the device‑load field to capture both the device’s internal queueing pressure and temporary throughput reductions, while disabling the egress‑port backpressure signal. The backpressure information is conveyed through telemetry messages and is treated as a congestion signal for CXL switches, since switches expose no explicit congestion notifications.


\noindent
{\bf Transmission Capacity Estimation.}
Due to the internal complexity of CXL hardware components, the transmission capacity is difficult to measure and may change dynamically. Inspired by the delay-based congestion control~\cite{tcpvegas-sigcomm94, timely-sigcomm15, dctcp-sigcomm10}, we maintain load factors \(l\) based on the congestion signals.
\begin{equation}
    l_{new} = (1 - g) \times l_{old} + g \times L
\end{equation}
\(L\) represents the percentage of responses marked as having moderate or severe overload in the device load field. For telemetry cases, the backpressure average percentage in the QoS response is directly used as \(L\). \(g\) is a configurable parameter. Using them, we then divide the operation region into the following categories and apply different scaling strategies:

\squishlist
    \item {\bf Congestion Avoidance}, occurring when fair share rate is increasing and \(l\) is relatively large. This means that some remote memory accesses have experienced high delays. We perform a multiplicative decrease and scale the reducing factor based on the load factor \(l\) (ALG\ref{alg:proto-alg} L22--23).
    
    \item {\bf Congestion Free}, where \(l < \delta\). This indicates that the rDIMM is able to deliver predefined bandwidth. Depending on how much idleness the channel has observed, our algorithm will add \(\lambda \times (\delta -h.l)\times K\) (ALG\ref{alg:proto-alg} L24--25).
        
\squishend

\subsection{Algorithm Extensions}
\label{subsec:proto-ext}

\noindent
\textbf{Weight Support.} \sys can be extended to support \texttt{mchannel}s with different weights \(w_c\), meaning that all congestion channels will receive a fair share rate of \(w_c\alpha_c\). The only modification required is to update Eq.~\ref{eq:time-alloc} accordingly.
\begin{equation}
\label{eq:weighted-time-alloc}
    T_R = \frac{\min(w_c\alpha_c, D_c)T_W}{D_c}
\end{equation}

\noindent
\textbf{HCSFQ Support.} Our algorithm achieves the fair share among \texttt{mchannel}s, equivalent to a single-layer HCSFQ.
However, in practice, an application may use multiple \texttt{mchannel}s (two-layer HCSFQ), and a tenant may run multiple applications (three-layer HCSFQ). Supporting HCSFQ is straightforward in \sys. In \(\S\)\ref{subsec:proto-alg}, we know (a) the transmission capacity of the hardware, which corresponds to the root node of the tree, and (b) memory access demand and access rate of the \texttt{mchannel}, which correspond to the leaf nodes of the tree. The aggregated demand/rate of the parent node is the sum of its children's (i.e., \(D_v = \sum D_u\)). By applying Eq.~\ref{eq:alpha-est} recursively from the root to the leaves, we can determine the fair share rate, \(\alpha\), for each layer. We then translate the \(\alpha\) of the last layer into the \texttt{mchannel} running time using Eq.~\ref{eq:time-alloc}. Note that the tree is used for all hardware components (adapters, switches, and DIMMs) as shown in our system model (\(\S\)\ref{subsec:channel-model}).

\subsection{Bound Analysis}
\label{subsec:proto-theoretical}

We present the following theorem to show that our algorithm provides performance bounds for applications. Consider a memory channel with a fixed fair share rate  \(\alpha_c\) and weight  \(w_c\) during a sampling window  \(K\). In the ideal case, the memory channel should not issue load/store traffic exceeding  \( w_c \alpha_c K \). We can prove that no matter how the application tries to game the system, our mechanism ensures that the bytes of load/store traffic from a memory channel are no more than: \(w_c\alpha_c K(1+2 \frac{T_W}{K})\),
where \(T_W\) is the scheduling window.

\noindent
\textbf{Proof.} Consider any interval \( K = [t', t'') \), and let \( h = \lfloor \frac{K}{T_W} \rfloor \). In an interval, there will be \(h\) full scheduling windows, at most one scheduling window before \( t' \), and at most one scheduling window after \( t'' \). Thus, the maximum number of bytes transmitted during \( K \) is \(\sum_{i=0}^{h+1} B_i\).
Based on Eq.~\ref{eq:weighted-time-alloc}, we have:
\begin{equation}
B_i \leq B_{fair} = \min(w_c \alpha_c, D_c) T_W \leq w_c \alpha_c T_W.
\end{equation}
Therefore, we can derive the bound as follows:
\begin{equation}
\begin{split}
\sum_{i=0}^{h+1} B_i \leq (h+2) w_c \alpha_c T_W 
\leq \left(\frac{K}{T_W} + 2\right) w_c \alpha_c T_W \\
= w_c \alpha_c K \left(1 + 2 \frac{T_W}{K}\right). \\
\end{split}
\end{equation}
%
This bound indicates that the application can issue at most \(2 \frac{T_W}{K}\) fractional more requests in the short run, ensuring that our algorithm achieves a max-min fair bandwidth allocation.

\section{Evaluation}
\label{sec:eval}

\subsection{Experimental Methodology}
\label{subsec:eval-exp-methodology}

{\bf Testbeds and Workloads.} We use the same experimental setup as $\S$\ref{subsec:chara-issues}. Our evaluations use a diverse set of memory-intensive workloads as prior studies~\cite{nomad-osdi24,memstrata-osdi24,soar-osdi25,colloid-sosp24}.
%
{\it (a). In-memory databases.} We run a hashtable-based in-memory key-value store, MICA~\cite{mica-nsdi14}, configured with 8-byte keys and 100-byte values, and we focus on its in-memory components: circular logs, lossy concurrent hash indexes, and bulk chaining. We also evaluate a B-tree–based database, Silo~\cite{silo-sosp13}, designed for low-latency transaction processing, with a 1KB value size. The client traffic is generated via YCSB~\cite{ycsb-socc}.
{\it (b). Graph analytics.} We deploy the GAP~\cite{gap-arxiv} benchmark suite, including several graph kernels: Breadth-First Search (BFS), Single-Source Shortest Paths (SSSP), PageRank (PR), Connected Components (CC), and Betweeness Centrality (BC), which use a Kronecker graph and a uniform random graph.
%
{\it (c). High-performance computing (HPC).} We also use SPEC CPU 2017~\cite{spec-cpu} and PARSEC~\cite{bienia2008parsec}, including scientific and numerical applications that stress memory access.

\noindent
{\bf Performance Metrics.} We report application throughput, latency, and performance slowdown. 
\sys obtains per-\textit{mchannel} bandwidth using performance counters (\S~\ref{subsec:mchannel-intf}) and aggregates them to derive both application-level and host-level memory bandwidth. We measure CXL memory access latency via CHA queues, i.e., dividing occupancy by the number of CXL requests, following the Colloid's approach~\cite{colloid-sosp24}.


\subsection{Application Performance over \sys}
\label{subsec:eval-app}

\begin{figure} [t!]
	\centering
	\subfloat[MICA latency.]{
	\begin{minipage}{0.24\textwidth}
	\includegraphics[width=\linewidth]{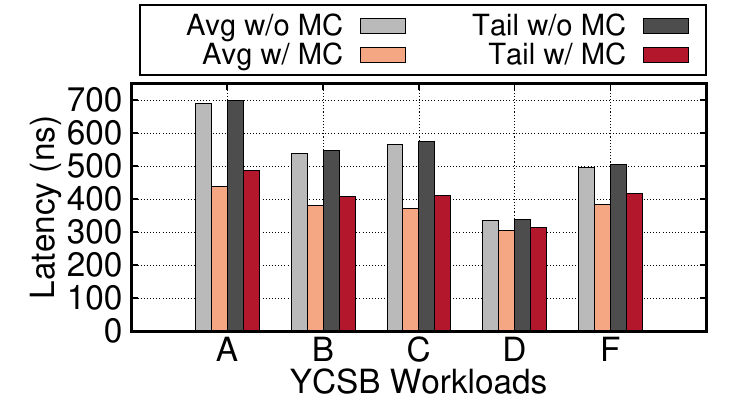}
        \end{minipage}\hfill
        }
	\subfloat[MICA throughput.]{
	\begin{minipage}{0.24\textwidth}
	\includegraphics[width=\linewidth]{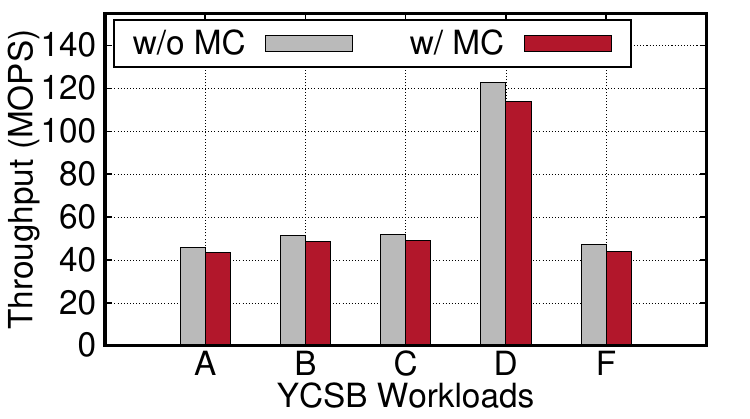}
        \end{minipage}\hfill
        }
        \\
        \subfloat[Silo latency.]{
	\begin{minipage}{0.24\textwidth}
	\includegraphics[width=\linewidth]{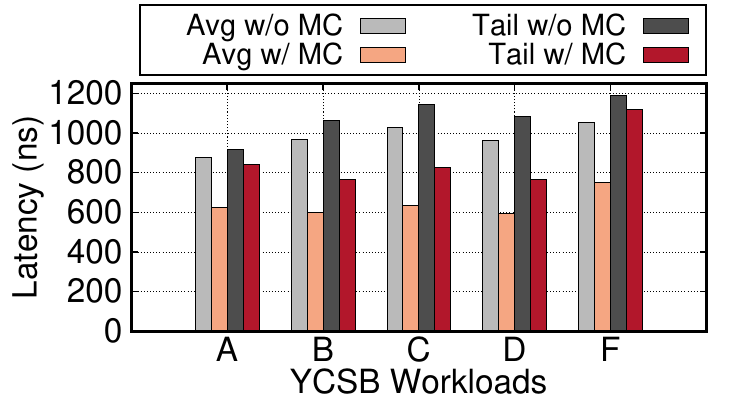}
        \end{minipage}\hfill
        }
        \subfloat[Silo throughput.]{
	\begin{minipage}{0.24\textwidth}
	\includegraphics[width=\linewidth]{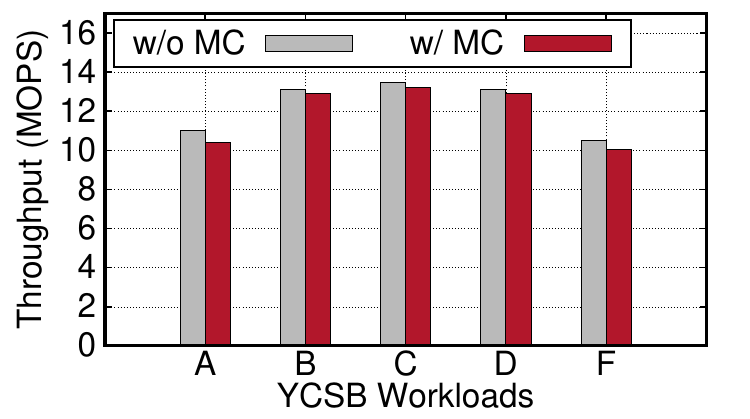}
        \end{minipage}\hfill
        }
        \vspace{-0.75\baselineskip}
        \caption{Throughput and latency of MICA and Silo over one $\times$8 adapter to the memory pool, comparing between w/ and w/o \sys scenarios. We use 32 threads in this experiment.}
        \vspace{-0.25\baselineskip}
        \label{fig:app-perf}
\end{figure}

\begin{figure} [t!]
	\centering
	\subfloat[MICA latency.]{
	\begin{minipage}{0.24\textwidth}
	\includegraphics[width=\linewidth]{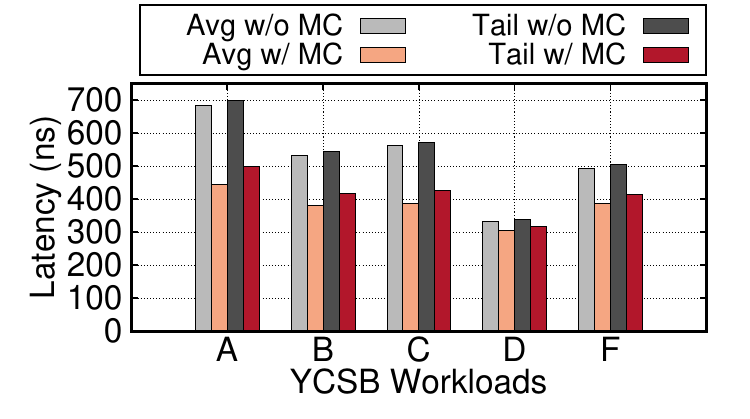}
        \end{minipage}\hfill
        }
	\subfloat[MICA throughput.]{
	\begin{minipage}{0.24\textwidth}
	\includegraphics[width=\linewidth]{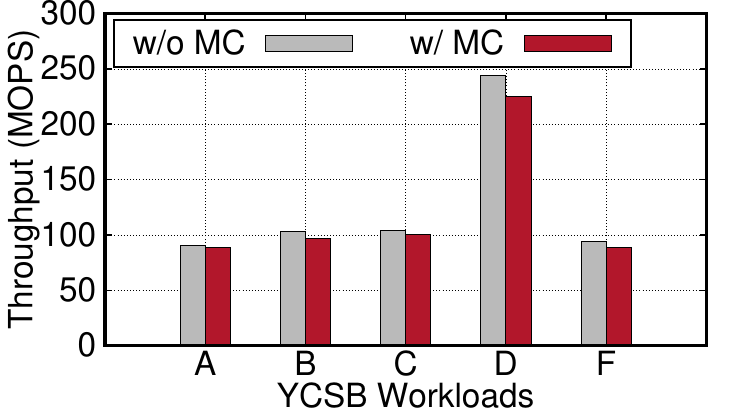}
        \end{minipage}\hfill
        }
        \\
        \subfloat[Silo latency.]{
	\begin{minipage}{0.24\textwidth}
	\includegraphics[width=\linewidth]{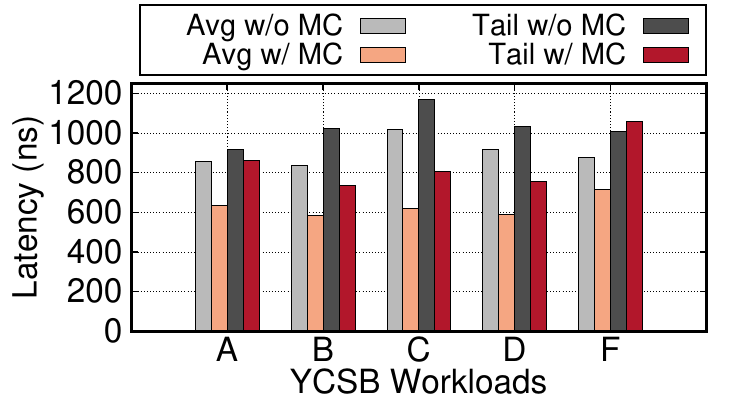}
        \end{minipage}\hfill
        }
        \subfloat[Silo throughput.]{
	\begin{minipage}{0.24\textwidth}
	\includegraphics[width=\linewidth]{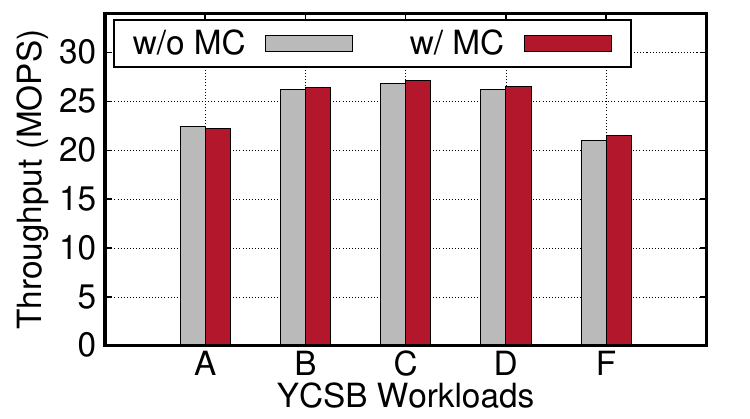}
        \end{minipage}\hfill
        }
        \vspace{-0.75\baselineskip}
        \caption{Throughput and latency of MICA and Silo over two $\times$8 adapters to the memory pool, comparing between w/ and w/o \sys scenarios. We use 64 threads in this experiment.}
        \vspace{-0.25\baselineskip}
        \label{fig:dual-app-perf}
\end{figure}

We first examine how much performance \sys delivers to applications when accessing the switched CXL memory pool. In this experiment, we use MICA and Silo and report the throughput and latency. With one $\times$8 CXL port, as shown in Figure~\ref{fig:app-perf}-a/c, applications show nearly no throughput degradation, and \sys can provide 19.6 GB/s bandwidth to the remote memory pool. However, \sys reduces the in-fabric congestion and yields latency savings. For example, as shown in Figure~\ref{fig:app-perf}-b/d, \sys reduces the average CXL memory access latency of Silo by 28.6\%/38.0\%/38.4\%/38.3\%/28.8\%, and tail latency by 8.7\%/28.0\%/27.7\%/29.3\%/5.8\% across five workloads compared with the cases when disabling \sys. MICA shows similar results. We then use two CXL adapters and run 64 threads. The average throughput over five applications reaches 225.3 MOPS under \textit{mchannel} (Figure~\ref{fig:dual-app-perf}), achieving 41.0GB/s CXL bandwidth to the remote memory pool. In terms of latency, similarly, \sys mitigates the fabric contention and achieves 23.9\% and 19.6\% lower average and tail latencies compared to the cases without \sys.



\subsection{Intra-Host Performance Isolation}
\label{subsec:eval-ha-contention}

\begin{figure} [t!]
	\centering
	\subfloat[Bwares Slowdown.]{
	\begin{minipage}{0.24\textwidth}
	\includegraphics[width=\linewidth]{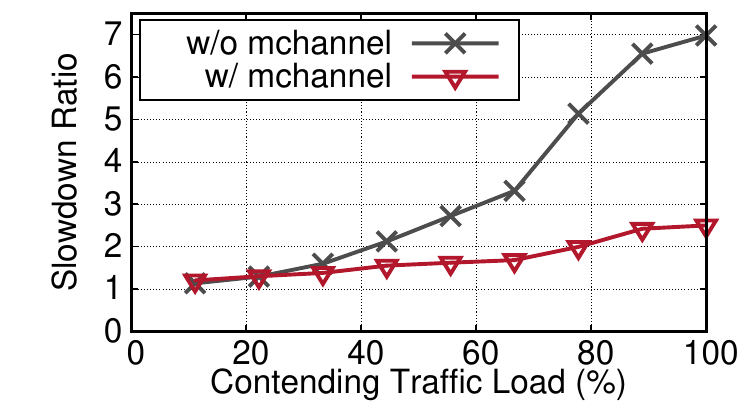}
        \end{minipage}\hfill
        }
	\subfloat[Roms Slowdown.]{
	\begin{minipage}{0.24\textwidth}
	\includegraphics[width=\linewidth]{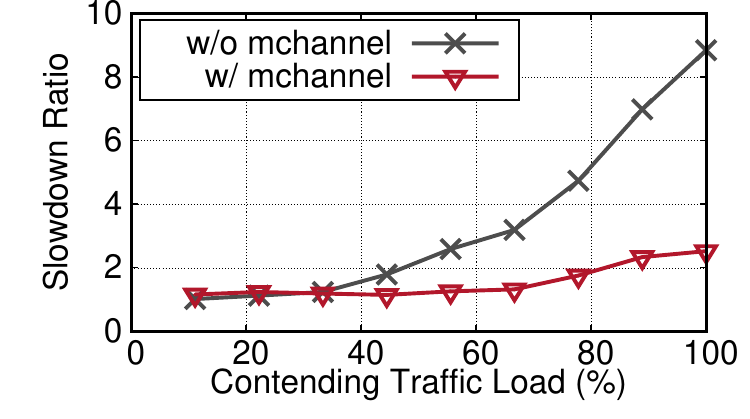}
        \end{minipage}\hfill
        }
        \\
    \subfloat[Bwares Bandwidth.]{
	\begin{minipage}{0.24\textwidth}
	\includegraphics[width=\linewidth]{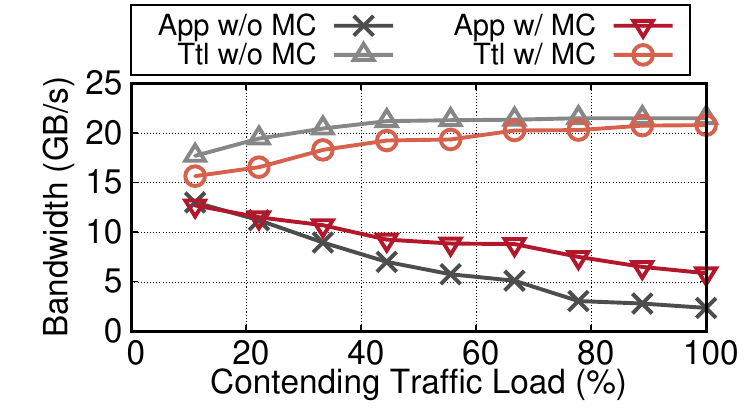}
        \end{minipage}\hfill
        }
	\subfloat[Roms Bandwidth.]{
	\begin{minipage}{0.24\textwidth}
	\includegraphics[width=\linewidth]{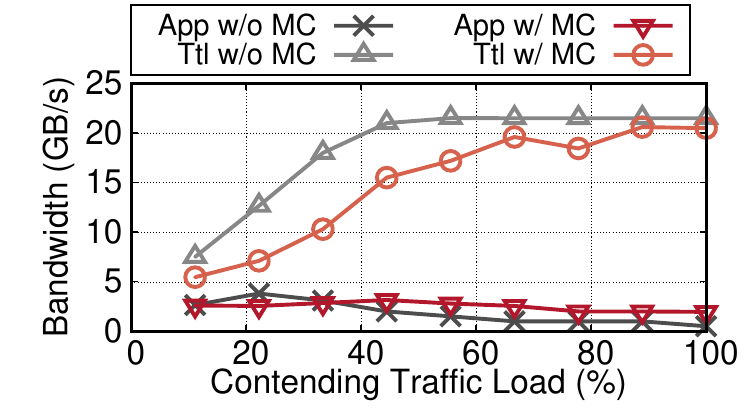}
        \end{minipage}\hfill
        }
        \\
        \subfloat[Bwares Latency.]{
	\begin{minipage}{0.24\textwidth}
	\includegraphics[width=\linewidth]{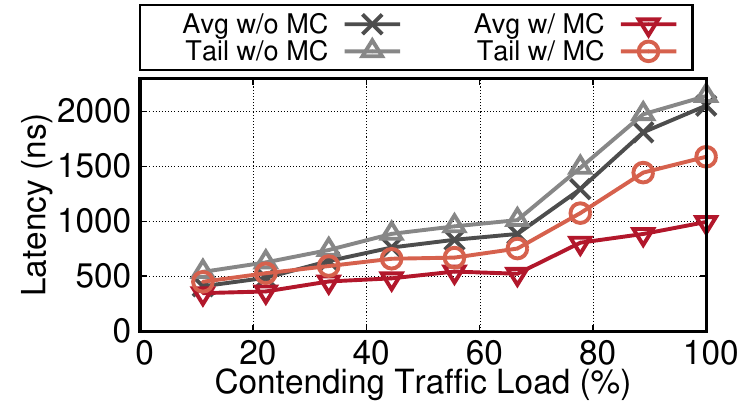}  
        \end{minipage}\hfill
        }
        \subfloat[Roms Latency.]{
	\begin{minipage}{0.24\textwidth}
	\includegraphics[width=\linewidth]{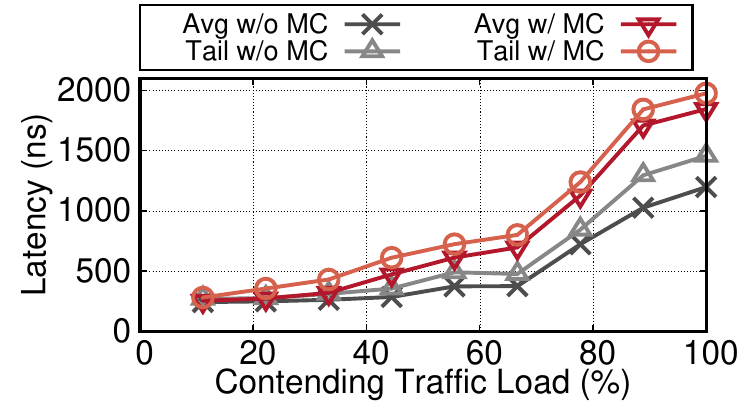}
        \end{minipage}\hfill
        }
        \vspace{-0.75\baselineskip}
        \caption{Application slowdown, CXL bandwidth, and latency when varying contending traffic under host adapter contention.}
        \vspace{-0.25\baselineskip}
        \label{fig:intra-host}
\end{figure}

We create intra-host contention by sharing the host adapter between an application and competing background traffic. 
We use Bwaves and Roms from CPU SPEC as our applications. As shown in Figure~\ref{fig:intra-host}-a/b, applications running without \sys experience $7.0\times$ and $8.8\times$ slowdowns as the contending traffic load increases to 100\% for bwaves and roms, respectively. In contrast, applications running under \sys only see a $2.5\times$ slowdown. This improvement occurs because \sys provides performance isolation among co-located channels through its transport algorithm (\S~\ref{subsec:proto-alg}). To better understand this effect, we further measure the CXL access bandwidth of the applications and the total bandwidth, as shown in Figure~\ref{fig:intra-host}-c/d. \sys allocates more bandwidth to applications based on max-min fairness. Under heavy contention (e.g., when the contending traffic load is $\geq 80\%$), \sys achieves an average of 96.1\% bandwidth utilization across two applications, while reducing average latency by 51.6\% and 35.1\% and tail latency by 25.8\% and 26.3\% for Bwaves and Roms, as shown in Figure~\ref{fig:intra-host}-e/f.

\subsection{Inter-Host Performance Isolation}
\label{subsec:eval-infabric}

\begin{figure} [t!]
	\centering
	\subfloat[BC Slowdown.]{
	\begin{minipage}{0.24\textwidth}
	\includegraphics[width=\linewidth]{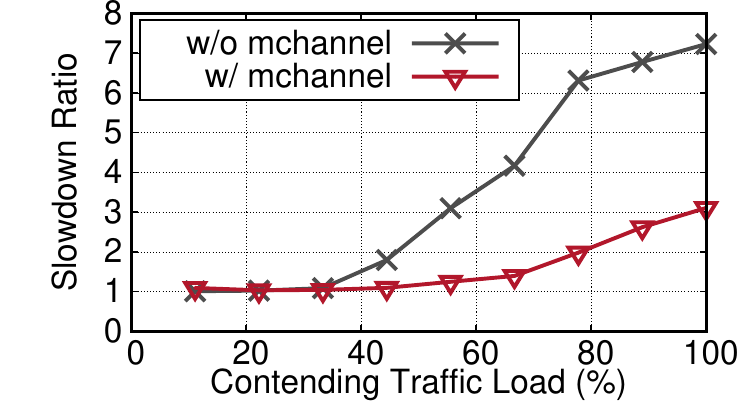}
        \end{minipage}\hfill
        }
	\subfloat[BC Bandwidth.]{
	\begin{minipage}{0.24\textwidth}
	\includegraphics[width=\linewidth]{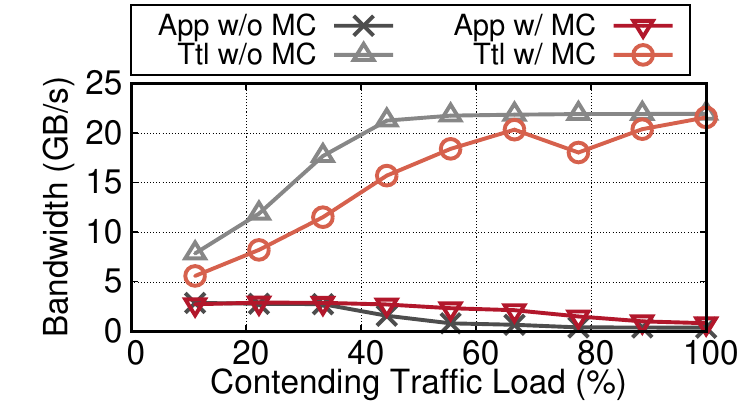}
        \end{minipage}\hfill
        }
        \\
        \subfloat[PR Slowdown.]{
	\begin{minipage}{0.24\textwidth}
	\includegraphics[width=\linewidth]{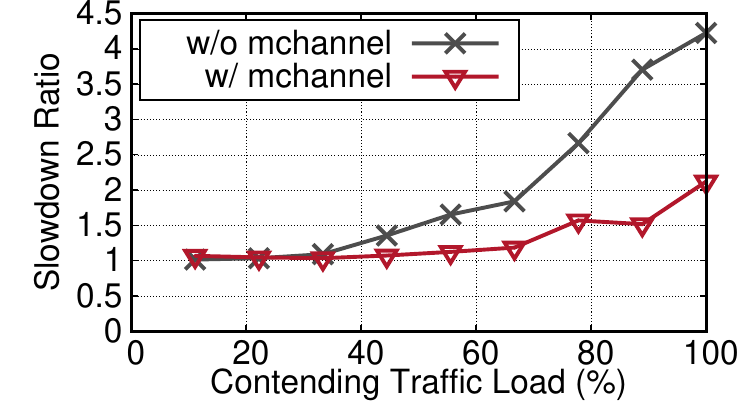}  
        \end{minipage}\hfill
        }
        \subfloat[PR Bandwidth.]{
	\begin{minipage}{0.24\textwidth}
	\includegraphics[width=\linewidth]{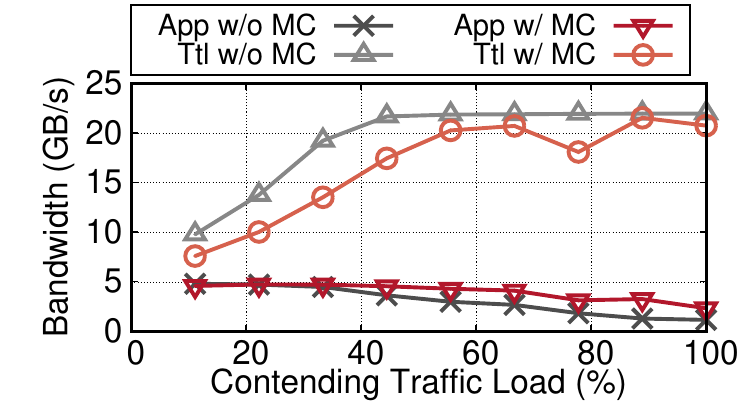}
        \end{minipage}\hfill
        }
        \vspace{-0.75\baselineskip}
        \caption{Application slowdown and CXL bandwidth when varying contending traffic under in-fabric congestion.}
        \vspace{-0.25\baselineskip}
        \label{fig:inter-host}
\end{figure}

We then set up in-fabric congestion as $\S$\ref{subsec:chara-issues} and evaluate how \sys mitigates the issue. We use two graph applications, i.e., Betweeness Centrality (BC) and PageRank (PR), in this experiment. As shown in Figure~\ref{fig:inter-host}-a/c, applications running without \sys experience $7.2\times$ and $4.2\times$ slowdowns as the contending traffic load increases to 100\% for BC and PR, respectively. In contrast, applications running under \sys only see 3.1$\times$ and 2.1$\times$ slowdown, respectively. The performance improvement comes from the fact that \sys allocates bandwidth across competing applications based on their performance requirements. Again, we measure the application CXL access bandwidth and the total bandwidth, as shown in Figure~\ref{fig:inter-host}-b/d. \sys allocates bandwidth in a max-min fairness manner at the CXL switching point. Under heavy contention (e.g., more than 80\% traffic load), \sys achieves an average of 97.2\% bandwidth utilization across two applications.

\subsection{Fairness}
\label{subsec:eval-fairness}

\begin{figure*} [t!]
	\centering
	\subfloat[Bwaves+SILO.]{
	\begin{minipage}{0.32\textwidth}
	\includegraphics[width=\linewidth]{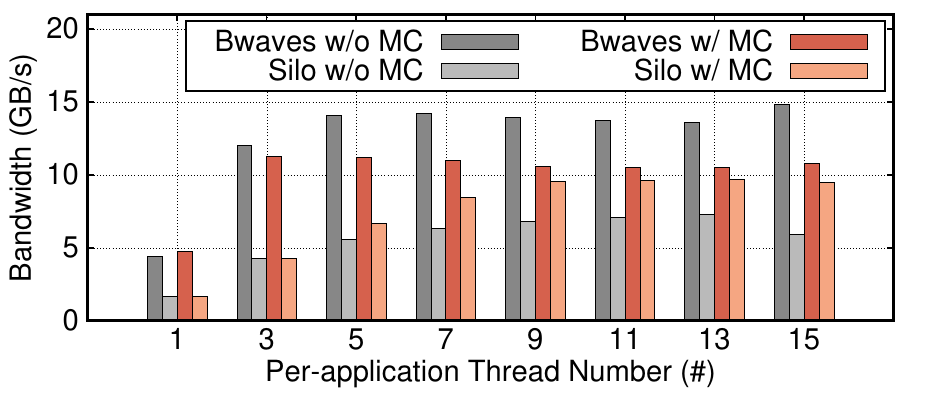}
        \end{minipage}\hfill
        }
	\subfloat[Bwaves+MICA.]{
	\begin{minipage}{0.32\textwidth}
	\includegraphics[width=\linewidth]{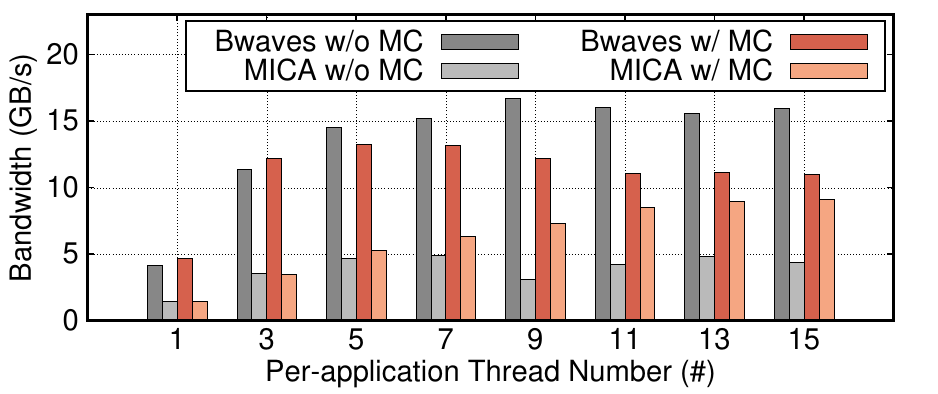}
        \end{minipage}\hfill
        }
        \subfloat[MICA+Silo.]{
	\begin{minipage}{0.32\textwidth}
	\includegraphics[width=\linewidth]{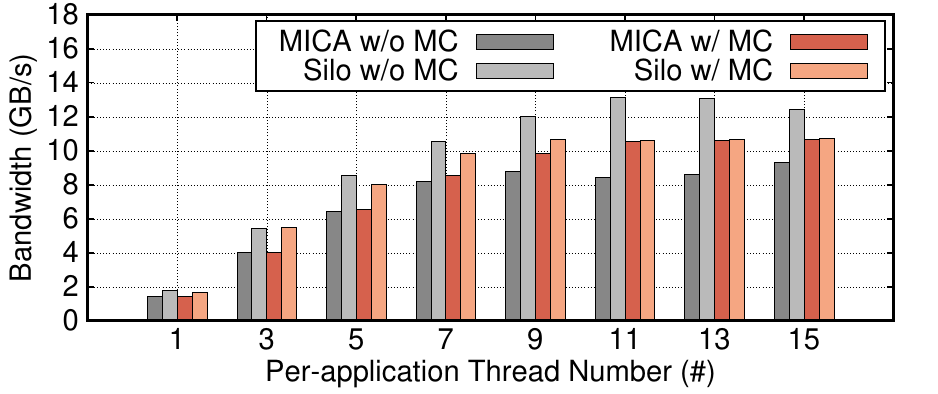}
        \end{minipage}\hfill
        }
        \vspace{-0.75\baselineskip}
        \caption{Compare the CXL memory bandwidth when running two applications with and without \sys support on a host. We gradually increase the thread number for both applications to create contention the memory fabric. }
        \vspace{-0.75\baselineskip}
        \label{fig:fairness}
\end{figure*}

\begin{figure*} [t!]
	\centering
	\subfloat[No Contention.]{
	\begin{minipage}{0.32\textwidth}
	\includegraphics[width=\linewidth]{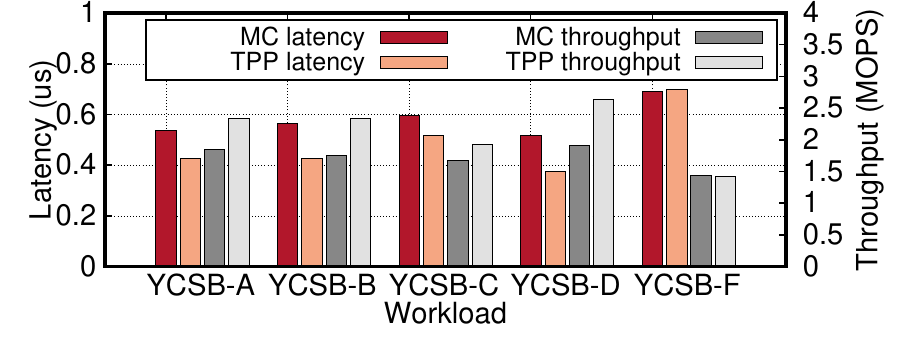}
        \end{minipage}\hfill
        }
	\subfloat[Moderate Contention.]{
	\begin{minipage}{0.32\textwidth}
	\includegraphics[width=\linewidth]{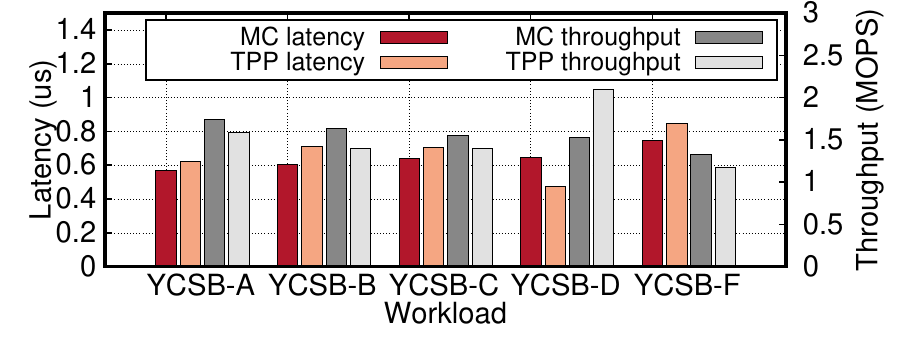}
        \end{minipage}\hfill
        }
        \subfloat[Heavy Contention.]{
	\begin{minipage}{0.32\textwidth}
	\includegraphics[width=\linewidth]{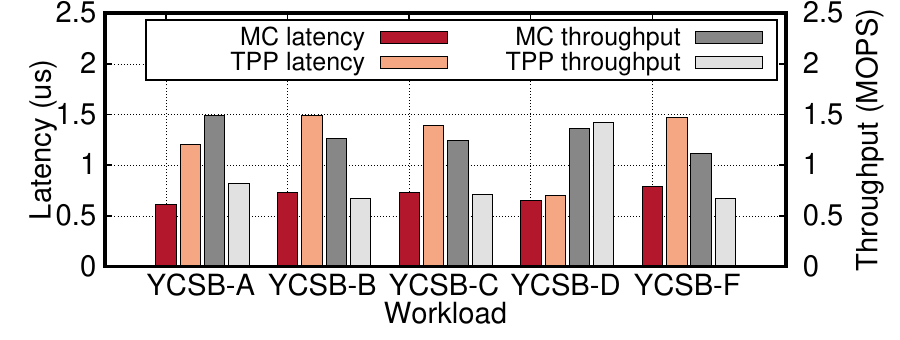}
        \end{minipage}\hfill
        }
        \vspace{-0.75\baselineskip}
        \caption{Latency and throughput for MICA running in \sys and TPP under five YCSB workloads. Y-1 axis is latency, and Y-2 axis is throughput. WS application is used to generate background traffic.}
        \vspace{-0.25\baselineskip}
        \label{fig:chann-app}
\end{figure*}

We achieve a fair bandwidth share by calculating the \texttt{mchannel} rate, and \sys enforces it via the rate control algorithm. To measure effectiveness, we co-locate pairs of SPEC CPU, MICA, and Silo applications, gradually increasing the number of threads for both applications to induce contention in the CXL fabric, and report the resulting memory bandwidth for each. As shown in Figure~\ref{fig:fairness}, under light contention (e.g., one thread), \sys matches the baseline by allocating bandwidth according to application demand. Under heavy contention (e.g., 15 threads), compared with the case without \sys, \sys significantly reduces the bandwidth gap between two applications: from 8.9GB/s to 1.3GB/s for Bwaves+Silo, from 11.5GB/s to 1.8GB/s for Bwaves+MICA, and from 3.1GB/s to 0.1GB/s for MICA+Silo. Our transport 
controls the application's remote access bandwidth, ensuring a fair share of hardware resources. 
We compare our runtime with TPP~\cite{tpp-asplos23} under different background traffic generated by the same SPEC application as above (Figure~\ref{fig:chann-app}). We run five types of YCSB workloads over MICA, with the local-to-remote memory capacity ratio set to eight. Without background traffic, TPP slightly outperforms our system by $1.2\times$ in terms of throughput and reduces latency by 16\% across all five workloads. This is because sufficient remote memory bandwidth is available to support page demotion and promotion. However, under moderate contention, our runtime matches the performance of TPP. Under heavy contention, our runtime achieves a $1.5\times$ throughput improvement and reduces latency by 43\%. This is due to the performance isolation capability provided by \sys.




\subsection{System Overheads}
\label{subsec:eval-overheads}

\sys incurs marginal overheads to the application.
Fair-share rate calculation and assignment are performed only once per scheduling window for all on-path devices, based on shared aggregated application demands and rates. Each core needs to read only one single performance counter via \textit{rdpmc}. The primary overhead introduced by \sys comes from POSIX timer timeout interruptions, used to suspend application execution. Based on our testing, this overhead amounts to 1.7\% when the scheduling window is set to 100 us, enabling fine-grained control of CXL memory access.

\subsection{Discussion}
\label{subsec:eval-discussion}

CXL ecosystems are under significant development. We believe that \sys can still be applied, but requires several changes. First, \sys currently advocates the device-internal load to indicate the traffic condition of each remote DIMM and uses end-to-end delay to estimate the fabric bandwidth capacity. With in-fabric explicit congestion notification,
our transport can be further improved. Second, \sys is evaluated at rack scale rather than at cluster scale. It will be interesting to explore how \sys operates in a scale-out CXL fabric with PBR. 
Third, the recent CXL specification introduces the CXL bundled port to support devices with higher bandwidth requirements. Since it enables logical aggregation of multiple CXL ports, \sys should account for the intra-bundle traffic condition and develop a new device-load estimation technique.

\section{Related Work}
\label{sec:related}

{\bf CXL Memory System.} 
Researchers have explored how to build CXL-based remote memory systems extensively~\cite{directcxl-atc22, tpp-asplos23, cxl-chara-micro23, melody-asplos25, fcc-hotos23, pathfinder-sigcomm25, pond-asplos23, nomad-osdi24, memstrata-osdi24}. 
For example, 
Pond~\cite{pond-asplos23} analyzes the memory stranding issue within Azure and proposes a CXL-based small pooled system architecture. 
Melody~\cite{melody-asplos25} develops a framework to characterize
CXL memory performance. 

\noindent
{\bf Load-Store Interconnect and Programmable Networks.} Load-store fabrics~\cite{cxl, rpcibench-nsdi24, ualink, nvlink} have gained great interest recently to build scale-up networks for resource disaggregation~\cite{gimbal-sigcomm21, dremel-sigmetrics22, leed-sigcomm23, ezns-osdi23, ezns-tos24, flint-nsdi25, ntprof-nsdi25, megastation-nsdi25, frsm-nsdi20, tapdb-asplos26}, but are mostly used in an exclusive deployment. As it moves to multi-tenant cases, similar transport techniques like \sys would be needed. We believe our past exploration on programmable networks~\cite{afq-nsdi18, calq-nsdi20, lognic-micro23, bf3charac-icnp24, rpcnic-hpca25, scr-nsdi25, sgiov-asplos26, pedal-nsdi26} and in-network computing~\cite{incbricks-asplos17, luo2017motivating, floem-osdi18, e3-atc19, ipipe-sigcomm19, liu2020building, clara-hotnets20, clara-sosp21, xenic-sosp21, ask-asplos23} can shed great light on.

\noindent
{\bf Congestion Control in Host Networks.} 
Researchers have built models to understand congestion happening within a single host~\cite{cai2021understanding, agarwal2023host, vuppalapati2024understanding, scn-hotnets25}.
HostCC~\cite{agarwal2023host} locates and identifies both host‑internal and network fabric congestion, significantly reducing host queueing and packet loss while improving throughput and tail latency.
Midhul Vuppalapati \emph{et al.} develop a domain‑by‑domain credit‑based flow control model to capture the subtle interplay that leads to latency inflation and host resource underutilization~\cite{vuppalapati2024understanding}.
\sys coordinates tenants in a load–store memory pooling setting and enables fair sharing of memory bandwidth.   

\section{Conclusion}
\label{sec:concl}

This paper presents \sys, a transport layer for switched CXL memory pooling. \sys introduces the \texttt{mchannel} abstraction for end-to-end fabric bandwidth management among competing memory streams and enables application-specific traffic control. Key to \sys is a Sender-Driven Fabric-Informed transport protocol--inspired by Core-Stateless Fair Queueing (CSFQ)--that admits just enough CXL requests to each \texttt{mchannel} based on the estimated $Core \leftrightarrow DIMM_{remote}$ bandwidth availability. Our evaluations demonstrate that \sys achieves high utilization, performance isolation, scalability, and multi-tenancy.


\section*{Acknowledgement}
We would like to thank the anonymous reviewers and our shepherd, Paolo Costa, for their comments and feedback. This work is supported in part by NSF grants CNS-2106199, CNS-2212192, CAREER-2339755, and Intel faculty awards.

\bibliographystyle{plain}
\bibliography{bibs/reference,bibs/ming}

@inproceedings{incbricks-asplos17, 
    author = {Liu, Ming and Luo, Liang and Nelson, Jacob and Ceze, Luis and Krishnamurthy, Arvind and Atreya, Kishore}, 
    title = {{IncBricks: Toward In-Network Computation with an In-Network Cache}}, 
    year = {2017}, 
    booktitle = {Proceedings of the Twenty-Second International Conference on Architectural Support for Programming Languages and Operating Systems (ASPLOS'17)},
    pages = {795–809}, 
}

@inproceedings{luo2017motivating,
  title={Motivating in-network aggregation for distributed deep neural network training},
  author={Luo, Liang and Liu, Ming and Nelson, Jacob and Ceze, Luis and Phanishayee, Amar and Krishnamurthy, Arvind},
  booktitle={Workshop on Approximate Computing Across the Stack},
  year={2017}
}

@inproceedings {afq-nsdi18,
    author = {Naveen Kr. Sharma and Ming Liu and Kishore Atreya and Arvind Krishnamurthy},
    title = {{Approximating Fair Queueing on Reconfigurable Switches}},
    booktitle = {15th USENIX Symposium on Networked Systems Design and Implementation (NSDI'18)},
    year = {2018},
    pages = {1--16},
}

@inproceedings {floem-osdi18,
    author = {Phitchaya Mangpo Phothilimthana and Ming Liu and Antoine Kaufmann and Simon Peter and Rastislav Bodik and Thomas Anderson},
    title = {{Floem: A Programming System for {NIC-Accelerated} Network Applications}},
    booktitle = {13th USENIX Symposium on Operating Systems Design and Implementation (OSDI'18)},
    year = {2018},
    pages = {663--679},
}

@inproceedings {e3-atc19,
    author = {Ming Liu and Simon Peter and Arvind Krishnamurthy and Phitchaya Mangpo Phothilimthana},
    title = {{E3: {Energy-Efficient} Microservices on {SmartNIC-Accelerated} Servers}},
    booktitle = {2019 USENIX Annual Technical Conference (USENIX ATC'19)},
    year = {2019},
    pages = {363--378},
}

@inproceedings{ipipe-sigcomm19, 
    author = {Liu, Ming and Cui, Tianyi and Schuh, Henry and Krishnamurthy, Arvind and Peter, Simon and Gupta, Karan}, 
    title = {{Offloading distributed applications onto smartNICs using iPipe}}, 
    year = {2019}, 
    booktitle = {Proceedings of the ACM Special Interest Group on Data Communication (SIGCOMM'19)}, 
    pages = {318–333}, 
}

@inproceedings {calq-nsdi20,
    author = {Naveen Kr. Sharma and Chenxingyu Zhao and Ming Liu and Pravein G Kannan and Changhoon Kim and Arvind Krishnamurthy and Anirudh Sivaraman},
    title = {{Programmable Calendar Queues for High-speed Packet Scheduling }},
    booktitle = {17th USENIX Symposium on Networked Systems Design and Implementation (NSDI'20)},
    year = {2020},
    pages = {685--699},
}

@inproceedings {frsm-nsdi20,
    author = {Ming Liu and Arvind Krishnamurthy and Harsha V. Madhyastha and Rishi Bhardwaj and Karan Gupta and Chinmay Kamat and Huapeng Yuan and Aditya Jaltade and Roger Liao and Pavan Konka and Anoop Jawahar},
    title = {{{Fine-Grained} Replicated State Machines for a Cluster Storage System }},
    booktitle = {17th USENIX Symposium on Networked Systems Design and Implementation (NSDI'20)},
    year = {2020},
    pages = {305--323},
}

@book{liu2020building,
  title={Building Distributed Systems Using Programmable Networks},
  author={Liu, Ming},
  year={2020},
  publisher={University of Washington}
}

@inproceedings{clara-hotnets20, 
    author = {Qiu, Yiming and Kang, Qiao and Liu, Ming and Chen, Ang}, 
    title = {{Clara: Performance Clarity for SmartNIC Offloading}}, 
    year = {2020}, 
    booktitle = {Proceedings of the 19th ACM Workshop on Hot Topics in Networks (HotNets'20)}, 
    pages = {16–22}, 
}

@inproceedings{gimbal-sigcomm21, 
    author = {Min, Jaehong and Liu, Ming and Chugh, Tapan and Zhao, Chenxingyu and Wei, Andrew and Doh, In Hwan and Krishnamurthy, Arvind}, 
    title = {{Gimbal: enabling multi-tenant storage disaggregation on SmartNIC JBOFs}}, year = {2021}, 
    booktitle = {Proceedings of the 2021 ACM SIGCOMM 2021 Conference (SIGCOMM'21)},
    pages = {106–122}, 
}

@inproceedings{xenic-sosp21, 
    author = {Schuh, Henry N. and Liang, Weihao and Liu, Ming and Nelson, Jacob and Krishnamurthy, Arvind}, 
    title = {{Xenic: SmartNIC-Accelerated Distributed Transactions}}, 
    year = {2021}, 
    booktitle = {Proceedings of the ACM SIGOPS 28th Symposium on Operating Systems Principles (SOSP'21)}, 
    pages = {740–755}, 
}

@inproceedings{clara-sosp21, 
    author = {Qiu, Yiming and Xing, Jiarong and Hsu, Kuo-Feng and Kang, Qiao and Liu, Ming and Narayana, Srinivas and Chen, Ang}, 
    title = {{Automated SmartNIC Offloading Insights for Network Functions}}, 
    year = {2021}, 
    booktitle = {Proceedings of the ACM SIGOPS 28th Symposium on Operating Systems Principles (SOSP'21)}, 
    pages = {772–787}, 
}

@article{dremel-sigmetrics22, 
    author = {Zhao, Chenxingyu and Chugh, Tapan and Min, Jaehong and Liu, Ming and Krishnamurthy, Arvind}, 
    title = {{Dremel: Adaptive Configuration Tuning of RocksDB KV-Store}}, 
    year = {2022}, 
    issue_date = {June 2022}, 
    volume = {6}, 
    number = {2},
    journal = {Proc. ACM Meas. Anal. Comput. Syst.}, 
    month = jun, 
    articleno = {37}, 
    numpages = {30}, 
}

@inproceedings{ask-asplos23, 
    author = {He, Yongchao and Wu, Wenfei and Le, Yanfang and Liu, Ming and Lao, ChonLam}, 
    title = {{A Generic Service to Provide In-Network Aggregation for Key-Value Streams}}, 
    year = {2023}, 
    booktitle = {Proceedings of the 28th ACM International Conference on Architectural Support for Programming Languages and Operating Systems (ASPLOS'23), Volume 2}, 
    pages = {33–47}, 
}

@inproceedings{fcc-hotos23, 
    author = {Liu, Ming}, 
    title = {{Fabric-Centric Computing}}, 
    year = {2023}, 
    booktitle = {Proceedings of the 19th Workshop on Hot Topics in Operating Systems (HotOS'23)}, 
    pages = {118–126}, 
}

@inproceedings {ezns-osdi23,
    author = {Jaehong Min and Chenxingyu Zhao and Ming Liu and Arvind Krishnamurthy},
    title = {{eZNS}: An Elastic Zoned Namespace for Commodity {ZNS} {SSDs}},
    booktitle = {17th USENIX Symposium on Operating Systems Design and Implementation (OSDI'23)},
    year = {2023},
    pages = {461--477},
}

@inproceedings{leed-sigcomm23, 
    author = {Guo, Zerui and Zhang, Hua and Zhao, Chenxingyu and Bai, Yuebin and Swift, Michael and Liu, Ming}, 
    title = {{LEED: A Low-Power, Fast Persistent Key-Value Store on SmartNIC JBOFs}},
    year = {2023},  
    booktitle = {Proceedings of the ACM SIGCOMM 2023 Conference (SIGCOMM'23)}, 
    pages = {1012–1027},
}

@inproceedings{lognic-micro23, 
    author = {Guo, Zerui and Lin, Jiaxin and Bai, Yuebin and Kim, Daehyeok and Swift, Michael and Akella, Aditya and Liu, Ming}, 
    title = {{LogNIC: A High-Level Performance Model for SmartNICs}}, 
    year = {2023}, 
    booktitle = {Proceedings of the 56th Annual IEEE/ACM International Symposium on Microarchitecture (MICRO'23)}, 
    pages = {916–929}, 
    numpages = {14}, 
}

@article{ezns-tos24, 
    author = {Min, Jaehong and Zhao, Chenxingyu and Liu, Ming and Krishnamurthy, Arvind}, 
    title = {{eZNS: Elastic Zoned Namespace for Enhanced Performance Isolation and Device Utilization}}, 
    year = {2024}, 
    issue_date = {August 2024}, 
    volume = {20}, 
    number = {3}, 
    journal = {ACM Trans. Storage}, 
    month = jun, 
    articleno = {16}, 
    numpages = {41}, 
}

@inproceedings{bf3charac-icnp24,
    title={{Demystifying datapath accelerator enhanced off-path smartnic}},
    author={Chen, Xuzheng and Zhang, Jie and Fu, Ting and Shen, Yifan and Ma, Shu and Qian, Kun and Zhu, Lingjun and Shi, Chao and Zhang, Yin and Liu, Ming and others},
    booktitle={2024 IEEE 32nd International Conference on Network Protocols (ICNP'24)},
    pages={1--12},
    year={2024},
}

@INPROCEEDINGS{rpcnic-hpca25,
    author={Zhang, Jie and Huang, Hongjing and Chen, Xuzheng and Li, Xiang and Zhao, Jieru and Liu, Ming and Wang, Zeke},
    booktitle={2025 IEEE International Symposium on High Performance Computer Architecture (HPCA'25)}, 
    title={{RpcNIC: Enabling Efficient Datacenter RPC Offloading on PCIe-attached SmartNICs}}, 
    year={2025},
    pages={1379-1394},
}

@inproceedings {flint-nsdi25,
    author = {Sheng Jiang and Ming Liu},
    title = {{Building an Elastic Block Storage over {EBOFs} Using Shadow Views}},
    booktitle = {22nd USENIX Symposium on Networked Systems Design and Implementation (NSDI'25)},
    year = {2025},
    pages = {1137--1153},
}

@inproceedings {ntprof-nsdi25,
    author = {Yuyuan Kang and Ming Liu},
    title = {{Understanding and Profiling {NVMe-over-TCP} Using ntprof}},
    booktitle = {22nd USENIX Symposium on Networked Systems Design and Implementation (NSDI'25)},
    year = {2025},
    pages = {1117--1136},
}

@inproceedings {megastation-nsdi25,
    author = {Xincheng Xie and Wentao Hou and Zerui Guo and Ming Liu},
    title = {Building Massive {MIMO} Baseband Processing on a {Single-Node} Supercomputer},
    booktitle = {22nd USENIX Symposium on Networked Systems Design and Implementation (NSDI'25)},
    year = {2025},
    pages = {1221--1242},
}

@inproceedings {scr-nsdi25,
    author = {Chenxingyu Zhao and Jaehong Min and Ming Liu and Arvind Krishnamurthy},
    title = {{{White-Boxing} {RDMA} with {Packet-Granular} Software Control}},
    booktitle = {22nd USENIX Symposium on Networked Systems Design and Implementation (NSDI'25)},
    year = {2025},
    pages = {427--449},
}

@inproceedings{pathfinder-sigcomm25, 
    author = {Li, Xiao and Guo, Zerui and Bai, Yuebin and Ketkar, Mehash and Willkinson, Hugh and Liu, Ming}, 
    title = {{Understanding and Profiling CXL.mem Using PathFinder}},
    year = {2025},  
    booktitle = {Proceedings of the ACM SIGCOMM 2025 Conference (SIGCOMM'25)}, 
}

@inproceedings{scn-hotnets25, 
    author = {An, Seunghyun and Oh, Joontaek and Liu, Ming}, 
    title = {{Server Chiplet Networking}}, 
    year = {2025}, 
    booktitle = {Proceedings of the 24th ACM Workshop on Hot Topics in Networks (HotNets'25)}, 
    pages = {289–299}, 
    numpages = {11},
}

@inproceedings{tapdb-asplos26,
  title = {{Understanding and Optimizing Database Pushdown on Disaggregated Storage}},
  author = {Zhang, Hua and Li, Xiao and Bai, Yuebin and Liu, Ming},
  booktitle = {Proceedings of the 31st ACM International Conference on Architectural Support for Programming Languages and Operating Systems, Volume 2},
  pages = {2141--2158},
  year = {2026}
}

@inproceedings{sgiov-asplos26,
  title = {{SG-IOV: Socket-Granular I/O Virtualization for SmartNIC-Based Container Networks}},
  author = {Zhao, Chenxingyu and Zhang, Hongtao and Min, Jaehong and Lin, Shengkai and Zhang, Wei and Zhang, Kaiyuan and Liu, Ming and Krishnamurthy, Arvind},
  booktitle = {Proceedings of the 31st ACM International Conference on Architectural Support for Programming Languages and Operating Systems, Volume 2},
  pages = {1727--1748},
  year = {2026}
}

@inproceedings {pedal-nsdi26,
    author = {Chendong Wang and Joontaek Oh and Ming Liu},
    title = {{Co-Designing} Traffic Control with {NVMe-oF} for Disaggregated Storage: A Comparative Study of Switched and Switchless {SAN} Architectures},
    booktitle = {23rd USENIX Symposium on Networked Systems Design and Implementation (NSDI'26)},
    year = {2026},
    pages = {2651--2669},
}

@MISC{intel-mlc,
  title = {{Intel Memory Latency Checker}},
  howpublished = {\url{https://www.intel.com/content/www/us/en/developer/articles/tool/intelr-memory-latency-checker.html}},
  year = {2025},
}

@MISC{cxl,
  title = {{Compute Express Link (CXL)}},
  howpublished = {\url{https://computeexpresslink.org}},
  year = {2025},
}

@MISC{pcie,
  title = {{PCI Express (Peripheral Component Interconnect Express)}},
  howpublished = {\url{https://pcisig.com}},
  year = {2025},
}

@MISC{cxl-spec,
  title = {{Compute Express Link (CXL) Specification}},
  howpublished = {\url{https://www.computeexpresslink.org/download-the-specification}},
  year = {2025},
}

@MISC{omega-fabric,
  title = {{IntelliProp's Omega Fabric}},
  howpublished = {\url{https://www.intelliprop.com/products-page}},
  year = {2025},
}

@MISC{samsung-cmmb,
  title = {{Samsung CXL Memory Module - Box (CMM-B)}},
  howpublished = {\url{https://semiconductor.samsung.com/news-events/tech-blog/cxl-memory-module-box-cmm-b/}},
  year = {2025},
}

@MISC{unifabrix-max,
  title = {{UnifabriX MAX}},
  howpublished = {\url{https://www.unifabrix.com/technology}},
  year = {2025},
}

@misc{h3-falcon,
  title={{The Falcon C5022}},
  howpublished = {\url{https://www.h3platform.com/product-detail/overview/35}},  
  year={2025}
}

@misc{cem-pcie,
  title={{An Introduction to Form Factors for PCI Express}},
  howpublished = {\url{https://pcisig.com/introduction-form-factors-pci-express}},  
  year={2025}
}

@misc{xconn-titan,
  title={{XConn Titan Evaluation Kit}},
  howpublished = {\url{https://www.xconn-tech.com/products}},  
  year={2025}
}

@MISC{smart-cxldimm,
  title = {{SMART CXL Memory Modules}},
  howpublished = {\url{https://www.smartm.com/product/list/cxl-memory?utm_source=CXL&utm_medium=Website&utm_term=CXL-Website-TR&utm_content=CXL-Website-Link&utm_campaign=CXL-Website}},
  year = {2025},
}

@MISC{micron-cxldimm,
  title = {{Micron Memory Expansion Using CXL}},
  howpublished = {\url{https://www.micron.com/products/memory/cxl-memory}},
  year = {2025},
}

@MISC{jedec-cxldimm,
  title = {{JEDEC Memory Module Reference Base Standard – for Compute Express Link (CXL)}},
  howpublished = {\url{https://www.jedec.org/standards-documents/docs/jesd317a}},
  year = {2025},
}

@inproceedings{hcsfq-nsdi21,
  title={Twenty years after: Hierarchical $\{$Core-Stateless$\}$ fair queueing},
  author={Yu, Zhuolong and Wu, Jingfeng and Braverman, Vladimir and Stoica, Ion and Jin, Xin},
  booktitle={18th USENIX Symposium on Networked Systems Design and Implementation (NSDI'21)},
  pages={29--45},
  year={2021}
}

@inproceedings{csfq-sigcomm98,
  title={Core-stateless fair queueing: Achieving approximately fair bandwidth allocations in high speed networks},
  author={Stoica, Ion and Shenker, Scott and Zhang, Hui},
  booktitle={Proceedings of the ACM SIGCOMM'98 conference on Applications, technologies, architectures, and protocols for computer communication},
  pages={118--130},
  year={1998}
}

@inproceedings{directcxl-atc22,
  title={{Direct access,$\{$High-Performance$\}$ memory disaggregation with $\{$DirectCXL$\}$}},
  author={Gouk, Donghyun and Lee, Sangwon and Kwon, Miryeong and Jung, Myoungsoo},
  booktitle={2022 USENIX Annual Technical Conference (USENIX ATC'22)},
  pages={287--294},
  year={2022}
}

@inproceedings{pond-asplos23, 
    author = {Li, Huaicheng and Berger, Daniel S. and Hsu, Lisa and Ernst, Daniel and Zardoshti, Pantea and Novakovic, Stanko and Shah, Monish and Rajadnya, Samir and Lee, Scott and Agarwal, Ishwar and Hill, Mark D. and Fontoura, Marcus and Bianchini, Ricardo}, 
    title = {{Pond: CXL-Based Memory Pooling Systems for Cloud Platforms}}, 
    year = {2023}, 
    booktitle = {Proceedings of the 28th ACM International Conference on Architectural Support for Programming Languages and Operating Systems (ASPLOS'23), Volume 2},
    pages = {574–587}, 
}

@inproceedings{tpp-asplos23, 
    author = {Maruf, Hasan Al and Wang, Hao and Dhanotia, Abhishek and Weiner, Johannes and Agarwal, Niket and Bhattacharya, Pallab and Petersen, Chris and Chowdhury, Mosharaf and Kanaujia, Shobhit and Chauhan, Prakash}, 
    title = {{TPP: Transparent Page Placement for CXL-Enabled Tiered-Memory}}, 
    year = {2023}, 
    booktitle = {Proceedings of the 28th ACM International Conference on Architectural Support for Programming Languages and Operating Systems (ASPLOS'23), Volume 3},
    pages = {742–755},
}

@inproceedings{cxlshm-sosp23,
  title={{Partial failure resilient memory management system for (cxl-based) distributed shared memory}},
  author={Zhang, Mingxing and Ma, Teng and Hua, Jinqi and Liu, Zheng and Chen, Kang and Ding, Ning and Du, Fan and Jiang, Jinlei and Ma, Tao and Wu, Yongwei},
  booktitle={Proceedings of the 29th Symposium on Operating Systems Principles (SOSP'23)},
  pages={658--674},
  year={2023}
}

@inproceedings{nomad-osdi24,
  title={{Nomad:$\{$Non-Exclusive$\}$ Memory Tiering via Transactional Page Migration}},
  author={Xiang, Lingfeng and Lin, Zhen and Deng, Weishu and Lu, Hui and Rao, Jia and Yuan, Yifan and Wang, Ren},
  booktitle={18th USENIX Symposium on Operating Systems Design and Implementation (OSDI'24)},
  pages={19--35},
  year={2024}
}

@inproceedings{dctcp-sigcomm10,
  title={Data center tcp (dctcp)},
  author={Alizadeh, Mohammad and Greenberg, Albert and Maltz, David A and Padhye, Jitendra and Patel, Parveen and Prabhakar, Balaji and Sengupta, Sudipta and Sridharan, Murari},
  booktitle={Proceedings of the ACM SIGCOMM 2010 Conference},
  pages={63--74},
  year={2010}
}

@inproceedings{cfc-sigcomm94,
  title={{Credit-based flow control for ATM networks: Credit update protocol, adaptive credit allocation and statistical multiplexing}},
  author={Kung, HT and Blackwell, Trevor and Chapman, Alan},
  booktitle={Proceedings of the conference on Communications architectures, protocols and applications},
  year={1994}
}

@article{cfc-network95,
  title={Credit-based flow control for ATM networks},
  author={Kung, NT and Morris, Robert},
  journal={IEEE network},
  volume={9},
  number={2},
  pages={40--48},
  year={1995},
}

@inproceedings{cxl-against-hotnets23, 
    author = {Levis, Philip and Lin, Kun and Tai, Amy}, 
    title = {{A Case Against CXL Memory Pooling}}, 
    year = {2023}, 
    booktitle = {Proceedings of the 22nd ACM Workshop on Hot Topics in Networks (HotNets'23)}, 
    pages = {18–24}, 
}

@article{cxlintro-arxiv23,
  title={{An introduction to the compute express link (cxl) interconnect}},
  author={Sharma, Debendra Das and Blankenship, Robert and Berger, Daniel S},
  journal={arXiv preprint arXiv:2306.11227},
  year={2023}
}

@inproceedings{cxl-chara-micro23, 
    author = {Sun, Yan and Yuan, Yifan and Yu, Zeduo and Kuper, Reese and Song, Chihun and Huang, Jinghan and Ji, Houxiang and Agarwal, Siddharth and Lou, Jiaqi and Jeong, Ipoom and Wang, Ren and Ahn, Jung Ho and Xu, Tianyin and Kim, Nam Sung}, 
    title = {{Demystifying CXL Memory with Genuine CXL-Ready Systems and Devices}}, 
    year = {2023}, 
    booktitle = {Proceedings of the 56th Annual IEEE/ACM International Symposium on Microarchitecture (MICRO'23)}, 
    pages = {105–121}, 
}

@MISC{pchase,
  title = {{pChase: A Pointer Chasing Benchmark}},
  howpublished = {\url{https://github.com/maleadt/pChase}},
  year = {2025},
}

@INPROCEEDINGS{intel-sapphire-isscc22,
  author={Nassif, Nevine and Munch, Ashley O. and Molnar, Carleton L. and Pasdast, Gerald and Lyer, Sitaraman V. and Yang, Zibing and Mendoza, Oscar and Huddart, Mark and Venkataraman, Srikrishnan and Kandula, Sireesha and Marom, Rafi and Kern, Alexandra M. and Bowhill, Bill and Mulvihill, David R. and Nimmagadda, Srikanth and Kalidindi, Varma and Krause, Jonathan and Haq, Mohammad M. and Sharma, Roopali and Duda, Kevin},
  booktitle={2022 IEEE International Solid-State Circuits Conference (ISSCC'22)}, 
  title={{Sapphire Rapids: The Next-Generation Intel Xeon Scalable Processor}}, 
  year={2022},
  volume={65},
  pages={44-46},
}

@inproceedings {rpcibench-nsdi24,
    author = {Wentao Hou and Jie Zhang and Zeke Wang and Ming Liu},
    title = {{Understanding Routable {PCIe} Performance for Composable Infrastructures}},
    booktitle = {21st USENIX Symposium on Networked Systems Design and Implementation (NSDI'24)},
    year = {2024},
}

@inproceedings{rmt-sigcomm13, 
    author = {Bosshart, Pat and Gibb, Glen and Kim, Hun-Seok and Varghese, George and McKeown, Nick and Izzard, Martin and Mujica, Fernando and Horowitz, Mark}, 
    title = {{Forwarding metamorphosis: fast programmable match-action processing in hardware for SDN}}, 
    year = {2013}, 
    booktitle = {Proceedings of the ACM SIGCOMM 2013 Conference on SIGCOMM}, 
    pages = {99–110}, 
    numpages = {12}, 
}

@inproceedings{tcpvegas-sigcomm94,
  title={{TCP Vegas: New techniques for congestion detection and avoidance}},
  author={Brakmo, Lawrence S and O'malley, Sean W and Peterson, Larry L},
  booktitle={Proceedings of the conference on Communications architectures, protocols and applications},
  pages={24--35},
  year={1994}
}

@inproceedings{timely-sigcomm15, 
    author = {Mittal, Radhika and Lam, Vinh The and Dukkipati, Nandita and Blem, Emily and Wassel, Hassan and Ghobadi, Monia and Vahdat, Amin and Wang, Yaogong and Wetherall, David and Zats, David}, 
    title = {{TIMELY: RTT-based Congestion Control for the Datacenter}}, 
    year = {2015}, 
    booktitle = {Proceedings of the 2015 ACM Conference on Special Interest Group on Data Communication (SIGCOMM'15)}, 
    pages = {537–550}, 
    numpages = {14}, 
}

@misc{asteralab-leo,
  title={{The Leo CXL™ Memory Connectivity Platform}},
  howpublished = {\url{https://www.asteralabs.com/products/cxl-memory-platform/leo-cxl-memory-connectivity-platform/}},  
  year={2025}
}

@misc{intel-agilex-dev-kit,
  title={{The Intel® Agilex™ 7 FPGA I-Series Development Kit}},
  howpublished = {\url{https://www.intel.com/content/www/us/en/products/details/fpga/development-kits/agilex/i-series/dev-agi027.html}},  
  year={2025}
}

@inproceedings{workstealing-ppopp13, 
    author = {L\^{e}, Nhat Minh and Pop, Antoniu and Cohen, Albert and Zappa Nardelli, Francesco}, 
    title = {{Correct and Efficient Work-Stealing for Weak Memory Models}}, 
    year = {2013},
    booktitle = {Proceedings of the 18th ACM SIGPLAN Symposium on Principles and Practice of Parallel Programming}, 
}

@inproceedings {mica-nsdi14,
    author = {Hyeontaek Lim and Dongsu Han and David G. Andersen and Michael Kaminsky},
    title = {{MICA}: A Holistic Approach to Fast {In-Memory} {Key-Value} Storage},
    booktitle = {11th USENIX Symposium on Networked Systems Design and Implementation (NSDI 14)},
    year = {2014},
}

@misc{gap-arxiv,
      title={{The GAP Benchmark Suite}}, 
      author={Scott Beamer and Krste Asanović and David Patterson},
      year={2017},
}

@inproceedings{ycsb-socc,
  title={{Benchmarking cloud serving systems with YCSB}},
  author={Cooper, Brian F and Silberstein, Adam and Tam, Erwin and Ramakrishnan, Raghu and Sears, Russell},
  booktitle={Proceedings of the 1st ACM symposium on Cloud computing},
  pages={143--154},
  year={2010}
}

@inproceedings{trie-accs07,
  title={{HAT-trie: a cache-conscious trie-based data structure for strings}},
  author={Askitis, Nikolas and Sinha, Ranjan},
  booktitle={Proceedings of the thirtieth Australasian conference on Computer science-Volume 62},
  pages={97--105},
  year={2007}
}

@inproceedings {memstrata-osdi24,
    author = {Yuhong Zhong and Daniel S. Berger and Carl Waldspurger and Ryan Wee and Ishwar Agarwal and Rajat Agarwal and Frank Hady and Karthik Kumar and Mark D. Hill and Mosharaf Chowdhury and Asaf Cidon},
    title = {{Managing Memory Tiers with {CXL} in Virtualized Environments}},
    booktitle = {18th USENIX Symposium on Operating Systems Design and Implementation (OSDI'24)},
    year = {2024},
    pages = {37--56},
}

@inproceedings{spa-asplos25,
  title={Systematic cxl memory characterization and performance analysis at scale},
  author={Liu, Jinshu and Hadian, Hamid and Wang, Yuyue and Berger, Daniel S and Nguyen, Marie and Jian, Xun and Noh, Sam H and Li, Huaicheng},
  booktitle={Proceedings of the 30th ACM International Conference on Architectural Support for Programming Languages and Operating Systems, Volume 2},
  pages={1203--1217},
  year={2025}
}

@inproceedings{collid-sosp24,
  title={{Tiered Memory Management: Access Latency is the Key!}},
  author={Vuppalapati, Midhul and Agarwal, Rachit},
  booktitle={Proceedings of the ACM SIGOPS 30th Symposium on Operating Systems Principles},
  pages={79--94},
  year={2024}
}

@inproceedings{aol-osdi25,
  title={{Tiered Memory Management Beyond Hotness}},
  author={Liu, Jinshu and Hadian, Hamid and Xu, Hanchen and Li, Huaicheng},
  booktitle={19th USENIX Symposium on Operating Systems Design and Implementation (OSDI'25)},
  pages={731--747},
  year={2025}
}

@inproceedings{hybridtier-asplos25,
  title={{HybridTier: an Adaptive and Lightweight CXL-Memory Tiering System}},
  author={Song, Kevin and Yang, Jiacheng and Wang, Zixuan and Zhao, Jishen and Liu, Sihang and Pekhimenko, Gennady},
  booktitle={Proceedings of the 30th ACM International Conference on Architectural Support for Programming Languages and Operating Systems, Volume 3},
  pages={112--128},
  year={2025}
}

@inproceedings{melody-asplos25,
  title={{Systematic cxl memory characterization and performance analysis at scale}},
  author={Liu, Jinshu and Hadian, Hamid and Wang, Yuyue and Berger, Daniel S and Nguyen, Marie and Jian, Xun and Noh, Sam H and Li, Huaicheng},
  booktitle={Proceedings of the 30th ACM International Conference on Architectural Support for Programming Languages and Operating Systems, Volume 2},
  pages={1203--1217},
  year={2025}
}

@incollection{hpcc-sigcomm19,
  title={{HPCC: High precision congestion control}},
  author={Li, Yuliang and Miao, Rui and Liu, Hongqiang Harry and Zhuang, Yan and Feng, Fei and Tang, Lingbo and Cao, Zheng and Zhang, Ming and Kelly, Frank and Alizadeh, Mohammad and others},
  booktitle={Proceedings of the ACM special interest group on data communication (SIGCOMM'19)},
  pages={44--58},
  year={2019}
}

@inproceedings{aeolus-sigcomm20,
  title={{Aeolus: A building block for proactive transport in datacenters}},
  author={Hu, Shuihai and Bai, Wei and Zeng, Gaoxiong and Wang, Zilong and Qiao, Baochen and Chen, Kai and Tan, Kun and Wang, Yi},
  booktitle={Proceedings of the Annual conference of the ACM Special Interest Group on Data Communication on the applications, technologies, architectures, and protocols for computer communication (SIGCOMM'20)},
  pages={422--434},
  year={2020}
}

@inproceedings{abm-sigcomm22,
  title={{ABM: Active buffer management in datacenters}},
  author={Addanki, Vamsi and Apostolaki, Maria and Ghobadi, Manya and Schmid, Stefan and Vanbever, Laurent},
  booktitle={Proceedings of the ACM SIGCOMM 2022 Conference (SIGCOMM'22)},
  pages={36--52},
  year={2022}
}

@inproceedings{harmony-nsdi24,
  title={Harmony: A congestion-free datacenter architecture},
  author={Agarwal, Saksham and Cai, Qizhe and Agarwal, Rachit and Shmoys, David and Vahdat, Amin},
  booktitle={21st USENIX Symposium on Networked Systems Design and Implementation (NSDI'24)},
  pages={329--343},
  year={2024}
}

@inproceedings{dcqcn-sigcomm15, 
    author = {Zhu, Yibo and Eran, Haggai and Firestone, Daniel and Guo, Chuanxiong and Lipshteyn, Marina and Liron, Yehonatan and Padhye, Jitendra and Raindel, Shachar and Yahia, Mohamad Haj and Zhang, Ming}, 
    title = {{Congestion Control for Large-Scale RDMA Deployments}}, 
    year = {2015},  
    booktitle = {Proceedings of the 2015 ACM Conference on Special Interest Group on Data Communication (SIGCOMM'15)}, 
    pages = {523–536}, 
}

@MISC{ci-wiki,
  title = {{Code Injection}},
  howpublished = {\url{https://en.wikipedia.org/wiki/Code_injection}},
  year = {2025},
}

@inproceedings{colloid-sosp24,
  title={{Tiered Memory Management: Access Latency is the Key!}},
  author={Vuppalapati, Midhul and Agarwal, Rachit},
  booktitle={Proceedings of the ACM SIGOPS 30th Symposium on Operating Systems Principles},
  pages={79--94},
  year={2024}
}

@inproceedings{silo-sosp13,
  title={Speedy transactions in multicore in-memory databases},
  author={Tu, Stephen and Zheng, Wenting and Kohler, Eddie and Liskov, Barbara and Madden, Samuel},
  booktitle={Proceedings of the Twenty-Fourth ACM Symposium on Operating Systems Principles},
  pages={18--32},
  year={2013}
}

@MISC{spec-cpu,
  title = {{SPEC CPU 2017}},
  howpublished = {\url{https://www.spec.org/cpu2017}},
  year = {2025},
}

@inproceedings{soar-osdi25,
  title={Tiered Memory Management Beyond Hotness},
  author={Liu, Jinshu and Hadian, Hamid and Xu, Hanchen and Li, Huaicheng},
  booktitle={19th USENIX Symposium on Operating Systems Design and Implementation (OSDI 25)},
  pages={731--747},
  year={2025}
}

@inproceedings{bienia2008parsec,
  title={The PARSEC benchmark suite: Characterization and architectural implications},
  author={Bienia, Christian and Kumar, Sanjeev and Singh, Jaswinder Pal and Li, Kai},
  booktitle={Proceedings of the 17th international conference on Parallel architectures and compilation techniques},
  pages={72--81},
  year={2008}
}

@inproceedings{cai2021understanding,
  title={Understanding host network stack overheads},
  author={Cai, Qizhe and Chaudhary, Shubham and Vuppalapati, Midhul and Hwang, Jaehyun and Agarwal, Rachit},
  booktitle={Proceedings of the 2021 ACM SIGCOMM 2021 Conference},
  pages={65--77},
  year={2021}
}

@inproceedings{agarwal2023host,
  title={Host congestion control},
  author={Agarwal, Saksham and Krishnamurthy, Arvind and Agarwal, Rachit},
  booktitle={Proceedings of the ACM SIGCOMM 2023 Conference},
  pages={275--287},
  year={2023}
}

@inproceedings{vuppalapati2024understanding,
  title={Understanding the host network},
  author={Vuppalapati, Midhul and Agarwal, Saksham and Schuh, Henry and Kasikci, Baris and Krishnamurthy, Arvind and Agarwal, Rachit},
  booktitle={Proceedings of the ACM SIGCOMM 2024 Conference},
  pages={581--594},
  year={2024}
}

@misc{ualink,
  title={{Ultra Accelerator Link (UALink)}},
  howpublished = {\url{https://www.ualinkconsortium.org}},  
  year={2026}
}

@misc{nvlink,
  title={{The NVIDIA NVLink}},
  howpublished = {\url{https://www.nvidia.com/en-us/data-center/nvlink/}},  
  year={2026}
}

\clearpage
\appendix
\section{Appendix}
\label{sec:appdix}

\subsection{\sys Adaptiveness}
\label{subsec:eval-util}

\begin{figure}[t]
    \includegraphics[width=\linewidth]{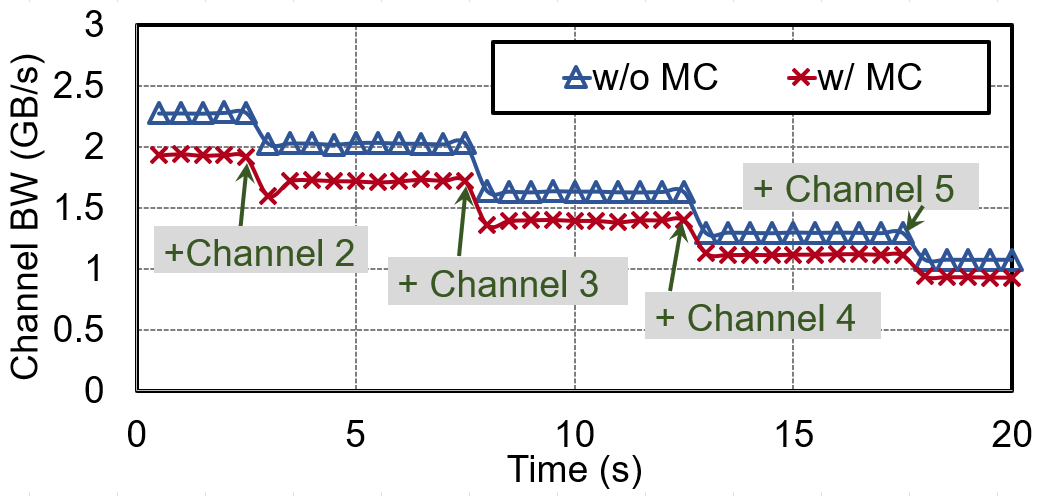}
    \vspace{-1.5\baselineskip}
    \caption{Channel bandwidth when adding four channels compared between with and without \sys cases.}
    \vspace{-0.25\baselineskip}
    \label{fig:chann-adapt}
\end{figure}

We further examine how the channel bandwidth adjusts when adding more \texttt{mchannel}s at runtime. We first run a synthetic workload with and without \sys. Then, at 2.5s, 7.5s, 12.5s, and 17.5s, we gradually add one more synthetic workload and measure the average channel bandwidth every half second. When a \texttt{mchannel} is added to the system, \sys updates the fair share rate and regulates the remote memory access rate (Figure~\ref{fig:chann-adapt}), causing a bandwidth drop at 3s, 8s, 13s, and 18s. Subsequently, the transport algorithm recalculates the hardware processing capacity based on congestion signals, increasing the average bandwidth.

\end{document}